\documentclass[english]{article}
\usepackage[T1]{fontenc}
\usepackage[latin9]{inputenc}
\usepackage[a4paper]{geometry}
\usepackage{babel}
\usepackage{array}
\usepackage{float}
\usepackage{url}
\usepackage{amsmath}
\usepackage{graphicx}
\usepackage{setspace}
\usepackage[unicode=true]{hyperref}
\usepackage{cite}

\makeatletter

\providecommand{\tabularnewline}{\\}

\DeclareMathOperator{\rmd}{d}

\makeatother

\begin{document}
\title{Order book model with herd behavior exhibiting long--range memory}
\author{Aleksejus Kononovicius, Julius Ruseckas}
\date{Institute of Theoretical Physics and Astronomy, Vilnius University}
\maketitle
\begin{abstract}
In this work, we propose an order book model with herd behavior. The
proposed model is built upon two distinct approaches: a recent empirical
study of the detailed order book records by Kanazawa \textit{et al.}
{[}\href{https://doi.org/10.1103/physrevlett.120.138301}{Phys. Rev. Lett. 120, 138301}{]}
and financial herd behavior model. Combining these approaches allows
us to propose a model that replicates the long-range memory of absolute
returns and trading activity. We compare the statistical properties
of the model against the empirical statistical properties of the Bitcoin
exchange rates and New York stock exchange tickers. We also show that
the fracture in the spectral density of the high--frequency absolute
return time series might be related to the mechanism of convergence
towards the equilibrium price.
\end{abstract}

\section{Introduction}

In the recent decades an increasing effort by social scientists, physicists
and broader interdisciplinary community has been applied to create
agent--based models (ABMs) of different social phenomena \cite{Helbing2010SciCu,Chakraborti2011RQUF1,Chakraborti2011RQUF2,Cristelli2012Fermi,Sornette2014,Abergel2017Springer,Chen2017FrontPhys,Flache2017JASSS}.
Notable part of these ABMs were created to explain various recurrent
anomalous statistical patterns, collectively referred to as stylized
facts, observed in the financial markets \cite{Chakraborti2011RQUF1,Chakraborti2011RQUF2,Cristelli2012Fermi,Chen2017FrontPhys}.
Financial ABMs vary in complexity usually trading plausibility for
analytical tractability \cite{Cristelli2012Fermi}. Some of the more
complex financial ABMs seem to be reasonably plausible, but they are
not analytically tractable. A well known example of such model is
the Lux--Marchesi model \cite{Lux1999Nature}. Hence it significantly
harder to understand their dynamics and to build upon them to further
improve their agreement with the empirical data.

In the recent few years we have developed a reasonably plausible yet
analytically tractable financial ABM \cite{Kononovicius2012PhysA,Gontis2014PlosOne}.
This financial ABM was derived from a widely recognized behavioral
model \cite{Kirman1993QJE}, which emphasizes imitation (herd) behavior
among socially interacting individuals. This financial ABM is able
to rather precisely fit the probability density functions (PDFs) and
power spectral densities (PSDs) of the empirical absolute return time
series. While the model has other desirable features, e.g., it scales
well with change in the time scale, it also has some drawbacks. First
of all it does not implement realistic trading strategies, though
most of the financial ABMs lack this feature \cite{Cristelli2012Fermi}.
But it is not the main issue we currently see. Our financial ABM relies
on a two assumptions, which are not fully and transparently justified.
We have assumed the presence of the omnipotent market maker, who is
able to clear the market instantaneously, (a rather common assumption
in many financial ABMs \cite{Chakraborti2011RQUF1,Chakraborti2011RQUF2,Cristelli2012Fermi})
and the presence of the exogenous noise. In our previous papers we
have speculated that exogenous noise is likely to originate from the
order book dynamics. Consequently it seems that including order book
dynamics in our financial ABM could potentially resolve these two
issues at once. Such approach has an additional perk, it would allow
us to consider trading activity time series as well.

Though the empirical order book data has become available at the same
time as the empirical high--frequency financial time series data,
in the 1980's, it took a longer time for detailed empirical studies
of the order book dynamics to be undertaken \cite{Biais1995,Gopikrishnan2000PRE,Bouchaud2002QF,Smith2003QF}.
Interestingly some of the observations were not as universal as the
stylized facts discovered in the empirical time series. There are
few recent empirical order book studies, which confirm some of the
earlier findings, but fail to confirm the other findings \cite{Alexandre2018}.
Similarly there is a variety of simple limit order book models (e.g.,
\cite{Bak1997PhysA,Maslov2000PhysA,Challet2001PhysA}), which are
not mutually compatible, but are able to reproduce some of the empirical
observations in the order book data. Some of the more recent order
book modeling approaches, such as \cite{Preis2006EPL,Cristelli2010EPJB,Abergel2013IJTAF,Donier2016JStatMech},
are more sophisticated and able to reproduce a variety of empirical
observations. The most recent approach by Kanazawa \textit{et al.}
\cite{Kanazawa2018PRL,Kanazawa2018PRE} combines empirical and theoretical
modeling approaches. Namely, Kanazawa \textit{et al.} have observed
the behavior of the high--frequency traders in a highly detailed
order book level data set. Based on the observations a microscopic
model inspired by the kinetic theory was proposed. Nevertheless most
of these models have not considered a detailed reproduction of the
stylized facts established for the time series data.

In this paper our goal is to introduce the order book mechanics, similar
to the ones introduced in \cite{Kanazawa2018PRL,Kanazawa2018PRE},
into our financial ABM \cite{Kononovicius2012PhysA,Gontis2014PlosOne}
thus producing a novel order book model with herd behavior, which
would be able to reproduce power law statistical properties of the
absolute return and trading activity time series. In Section~\ref{sec:review}
we will compare the order book modeling approach we will take here
against a few similar recently published approaches \cite{Kaizoji2015,Biondo2016JNTF,Biondo2018Eco,Biondo2018JEcoIntCo,Biondo2018SEF,Navarro2017Plos,Cocco2017JEIC,Cocco2019FI,Llacay2018CompMatOrgT}.
In Section~\ref{sec:kirman-model} we will briefly introduce herd
behavior model proposed by Kirman \cite{Kirman1993QJE}, which is
the basis of our agent--based approach. In Section~\ref{sec:financial-herd}
we will discuss two different financial market models: the previously
proposed herd behavior model with instantaneous clearing \cite{Kononovicius2012PhysA,Gontis2014PlosOne}
and the order book model with herd behavior. We will show that under
certain parameter values both of the models produce mostly identical
statistical properties. Further in Section~\ref{sec:empirical} we
will compare the order book model with herd behavior against the empirical
high--frequency Bitcoin and NYSE data. We will examine the sensitivity
of the model to parameter value changes in Section~\ref{sec:params}.
While we will provide concluding remarks and future outlooks in Section~\ref{sec:conclusions}.

\section{Review of the other similar order book modeling approaches\label{sec:review}}

Recently a few approaches similar to the one we will take here were
published \cite{Kaizoji2015,Biondo2016JNTF,Biondo2018Eco,Biondo2018JEcoIntCo,Biondo2018SEF,Navarro2017Plos,Cocco2017JEIC,Cocco2019FI,Llacay2018CompMatOrgT}.
While these approaches are different in many aspects, we will keep
our review brief by highlighting only the most important similarities
and differences between these approaches themselves and our approach
to be taken in the next sections of this paper. The review is summarized
in Table~\ref{tab:compare-approaches}.

\begin{table}
\caption{Comparison of the approaches reviewed in Section~\ref{sec:review}.
Differences between our approach and the reviewed approaches are highlighted
using \textit{italics}.\label{tab:compare-approaches}}

\centering{}%
\begin{tabular}{|c|p{3.5cm}|c|c|c|c|}
\hline 
\textbf{Approach} & \textbf{Agent types} & \textbf{Fixed/flexible} & \textbf{Strategies} & \textbf{Volume} & \textbf{Tractable}\tabularnewline
\hline 
\hline 
\cite{Kaizoji2015} & chartists, fundamentalists & \textit{fixed} & \textit{sound} & \textit{strategic} & partly analytically\tabularnewline
\hline 
\cite{Biondo2016JNTF} & chartists, fundamentalists & \textit{fixed} & simple & unit & \textit{only numerically}\tabularnewline
\hline 
\cite{Biondo2018JEcoIntCo,Biondo2018SEF,Biondo2018Eco} & chartists, fundamentalists & \textit{fixed} & \textit{sound} & \textit{random} & \textit{only numerically}\tabularnewline
\hline 
\cite{Navarro2017Plos} & chartists, fundamentalists & \textit{fixed} & \textit{realistic} & unit & \textit{only numerically}\tabularnewline
\hline 
\cite{Cocco2017JEIC,Cocco2019FI} & chartists,\textit{ random traders} & \textit{fixed} & simple & \textit{random} & \textit{only numerically}\tabularnewline
\hline 
\cite{Llacay2018CompMatOrgT} & chartists, fundamentalists & \textit{fixed} & \textit{realistic} & \textit{strategic} & \textit{only numerically}\tabularnewline
\hline 
Our & chartists, fundamentalists & flexible & simple & unit & possibly analytically\tabularnewline
\hline 
\end{tabular}
\end{table}

First of all it is important to note that all of the models under
our review are able to reproduce the main stylized facts of absolute
return time series. Some of the models are also able to reproduce
the features of price time series \cite{Kaizoji2015} or specific
peculiarities observed in the modeled markets \cite{Cocco2017JEIC,Cocco2019FI}.
Neither of the considered approaches reports results on the reproducibility
of the statistical properties of trading activity. Our goal will be
to reproduce PDFs and PSDs of both absolute return and trading activity
time series. Note that some of the approaches reviewed here report
auto--correlation functions of the return time series, while we prefer
to use PSDs. This makes no significant difference as PSD is directly
related to the auto--correlation function via the Wiener--Khinchin
theorem.

All of the considered models assume that there are two types of traders.
One type of traders relies on the internal market information. These
traders are usually called chartists. Another type of traders use
exogenous information. In most of the works exogenous information
is some kind of parametrization of the fundamental price, hence such
traders are called chartists. Among the considered works only Cocco
\textit{et al.} \cite{Cocco2017JEIC,Cocco2019FI} make a different
assumption. Namely, Cocco \textit{et al.} assume that exogenous information
is related to the objective needs of traders, which are unknown and
random, therefore it causes random trading. Our approach is similar
to the most of these works, we will assume that the market is populated
by the two types of agents: chartists and fundamentalists.

Unlike in these models in our approach agents are flexible, namely
they are able to switch their trading strategies. It is quite a puzzle
to explain why the models with fixed and flexible agent types are
able to reproduce the main stylized facts. Most likely it is because
these models rely on the heterogeneity of the agents within each type.
Whenever this intra--type heterogeneity breaks down, the agents act
similarly on the received information, and thus large fluctuations
emerge. In our approach large fluctuations also emerge when most agents
become chartists, i.e., when inter-type heterogeneity breaks down.
In all cases fluctuations will be the largest under heterogeneity
breaking.

The approaches also differ on how the trading strategies are specified.
Most realistic specification can be found in \cite{Navarro2017Plos,Llacay2018CompMatOrgT}
as the model proposed in these works utilize technical trading strategies
which are actually used for trading in the real markets. A bit less
practical, yet sound from an economic perspective, strategies can
be found in \cite{Kaizoji2015,Biondo2018JEcoIntCo,Biondo2018SEF,Biondo2018Eco}.
These strategies are economically sound because they allow the agents
to maximize their expected wealth, though these strategies are not
necessarily common among the traders in the real markets. While our
approach, and also \cite{Biondo2016JNTF,Cocco2017JEIC,Cocco2019FI},
is the simplest in this context. Namely, trading strategies in this
case are assumed to be highly stylized and do not necessarily guarantee
wealth gain for the agents. In some sense it could be said that the
agents under this approach have zero intelligence.

We could categorize the approaches to the trading strategies from
another point of view. The strategies also significantly differ on
how the order volumes are treated. We find three different possible
treatments: order volume is selected strategically in \cite{Kaizoji2015,Llacay2018CompMatOrgT}
(it is important to note that neither of these models uses order books),
order volume is drawn from random distribution in \cite{Cocco2017JEIC,Cocco2019FI,Biondo2018JEcoIntCo,Biondo2018SEF,Biondo2018Eco}
or it is set to be of unit size \cite{Biondo2016JNTF,Navarro2017Plos}.
In \cite{Kanazawa2018PRL,Kanazawa2018PRE}, on which we will based
our order book approach, order volumes are also assumed to be of unit
size, this assumption is backed up by empirical analysis, which shows
that around $80\%$ of completed transactions fill a unit of volume.

Although the differences in respect to trading strategies can be extremely
important when trying to build a model for economic policy making
\cite{Navarro2017Plos,Biondo2018Eco,Biondo2018JEcoIntCo,Biondo2018SEF,Llacay2018CompMatOrgT},
it does not help to make model more tractable, which is also a rather
desirable feature. All of these models, with the exception of \cite{Kaizoji2015},
are not analytically tractable and can be studied only via numerical
simulation. Thus only limited knowledge about their dynamics can be
obtained. Though our financial ABM seems to be comparatively lacking
in economic plausibility it makes up by being analytically tractable,
though one has to rely on Stochastic Calculus for that. As the model
in \cite{Kaizoji2015} our financial ABM lacks order book clearing
mechanism, but in this work we will introduce that in to our financial
ABM. As the order book modeling approach we will take in this work
is analytically tractable \cite{Kanazawa2018PRL,Kanazawa2018PRE},
we can expect that the combination of the both approaches would also
become analytically tractable at some point.

\section{Kirman's herd behavior model\label{sec:kirman-model}}

Let us start with discussion about Kirman's herd behavior model \cite{Kirman1993QJE},
which is the base upon we build financial market models in the following
sections. In the seminal paper Kirman shared an observation that social
scientists and behavioral biologists observe remarkably similar patterns
in rather distinct systems. In the experiments involving ants described
by Kirman, entomologists observed the emergence of asymmetry in a
symmetric experimental setup: despite having two identical food sources
available, majority of the ants in the ant colony preferred to forage
from a single food source at a time. Numerous references in Kirman's
paper suggest that humans also seem to prefer the more popular product
over the less popular despite both being of a similar quality.

To account for these empirical observations Kirman proposed a simple
probabilistic herd behavior model in which the probability for an
agent to switch to another state is proportional to the fraction of
agents in that state:
\begin{align}
p\left(X\rightarrow X+1\right) & =\left(N-X\right)\left(\sigma_{1}+h\frac{X}{N}\right)\Delta t,\\
p\left(X\rightarrow X-1\right) & =X\left[\sigma_{2}+h\frac{N-X}{N}\right]\Delta t,
\end{align}
where $X$ is a number of agents in the first state, $N$ -- a total
number of agents, $\sigma_{i}$ -- an idiosyncratic behavior parameter
(encodes preferences for the states), $h$ herd behavior parameter,
$\Delta t$ -- arbitrarily small time step. This formulation of the
herd behavior model is often referred to as ``local'' or ``extensive'',
because the fluctuations of $X$ quickly disappear as $N$ becomes
larger \cite{Alfarano2009Dyncon,Flache2011JCR,Kononovicius2014EPJB,Peralta2018VMNet}.
In other words $X$ rapidly converges to a certain value and remains
almost constant afterwards.

What we described above in the literature is often referred to as
the $N$-dependence problem \cite{Alfarano2009Dyncon,Flache2011JCR,Kononovicius2014EPJB}.
This problem can be circumvented by assuming that the probability
to switch is proportional to a total number of agents $X$:
\begin{align}
p\left(X\rightarrow X+1\right) & =\left(N-X\right)\left(\sigma_{1}+hX\right)\Delta t,\label{eq:nonextplus}\\
p\left(X\rightarrow X-1\right) & =X\left[\sigma_{2}+h\left(N-X\right)\right]\Delta t,\label{eq:nonextmin}
\end{align}
To contrast the previous formulation of the model, this formulation
of the herd behavior model is often referred to as the ``non-extensive''
or ``global'' formulation. In this formulation $X$ no longer converges
to a fixed value even in the limit of infinite $N$. This is desirable
feature to have in the financial market models as it is well-known
that the stylized facts hold for small and large markets alike \cite{Cont2001RQUF}.
Hence there is a variety of the financial ABMs, which were inspired
by the non-extensive formulation of the Kirman's model \cite{Lux1999Nature,Alfarano2005CompEco,Alfarano2008Dyncon,Alfi2009EPJB1,Alfi2009EPJB2,Kononovicius2012PhysA,Gontis2014PlosOne}.
There are also papers in the opinion dynamics, which claim that the
fluctuating nature of opinion change can be explained by assuming
the presence of collective peer-pressure instead of inter-personal
communication \cite{Flache2011JCR,Kononovicius2014EPJB,Sano2016,Kononovicius2017Complexity,Peralta2018VMNet}.

Let us take the infinite $N$ limit and introduce an almost continuous
state variable $x=\frac{X}{N}$. This allows us to rewrite the model
driven by Eqs.~(\ref{eq:nonextplus}) and (\ref{eq:nonextmin}) as
a stochastic differential equation \cite{Alfarano2005CompEco,Alfarano2008Dyncon,Kononovicius2012PhysA}:

\begin{equation}
\rmd x=h\left[\varepsilon_{1}\left(1-x\right)-\varepsilon_{2}x\right]\rmd t+\sqrt{2hx\left(1-x\right)}\rmd W,\label{eq:sdex}
\end{equation}
where $\varepsilon_{i}=\frac{\sigma_{i}}{h}$. From the Eq.~(\ref{eq:sdex})
it is straightforward to conclude that $x$ is Beta distributed, $x\sim\text{Beta}\left(\varepsilon_{1},\varepsilon_{2}\right)$.
If we would set $\varepsilon_{1}=\varepsilon_{2}$ and $\varepsilon_{1}<1$
then we would observe the same pattern entomologists did as the Beta
distribution is multi modal in this case.

One can derive a similar SDE, by taking finitely large $N$ limit
under extensive formulation of the model. Yet in this case SDE becomes
ordinary differential equation as with larger $N$ the diffusion term
becomes negligible and only the drift term remains:
\begin{align}
\rmd x & =h\left[\varepsilon_{1}\left(1-x\right)-\varepsilon_{2}x\right]\rmd t+\sqrt{\frac{2hx\left(1-x\right)}{N}}\rmd W\approx\nonumber \\
 & \approx h\left[\varepsilon_{1}\left(1-x\right)-\varepsilon_{2}x\right]\rmd t.
\end{align}
In a special case, $p\left(X\rightarrow X-1\right)=0$, one can also
derive the well-known Bass diffusion equation \cite{Kononovicius2012IntSys}:
\begin{equation}
\rmd x=\left(1-x\right)\left(\sigma_{1}+hx\right)\rmd t.
\end{equation}

\section{Herd behavior model in the context of the financial markets\label{sec:financial-herd}}

Kirman's herd behavior model is in a sense a generic model. In the
financial market context it would be natural for these states to represent
different trading strategies. In the agent--based modeling literature
one can find the most common is to consider interaction between agents
using fundamentalist and chartist trading strategies \cite{Cristelli2012Fermi,Abergel2017Springer,Llacay2018CompMatOrgT}.
It is worth to note a couple approaches which consider modeling market
sentiments instead \cite{Alfarano2008Dyncon,Chen2016CE,Lux2018JEDC}.
In our previous works \cite{Kononovicius2012PhysA,Gontis2014PlosOne}
we have used both of these approaches to build a highly sophisticated
ABM which was able to fit the empirical absolute return PDF and PSD
rather well. Here we will start with the fundamentalist-chartist model
under instantaneous clearing and later use it to build order book
model.

\subsection{Herd behavior model with instantaneous clearing}

Let us define the trading strategies in a rather stylized manner.
One could in general define more sophisticated strategies (as discussed
in \cite{Llacay2018CompMatOrgT}), but we would like to keep the model
compatible with our earlier works. This will allow us to use some
of the earlier analytically obtained results which would be impossible
while using a more realistic trading strategies.

In \cite{Kononovicius2012PhysA} we have assumed that excess demand
by chartist traders is conditioned on their mood $\xi\left(t\right)$:
\begin{equation}
D_{c}\left(t\right)=r_{0}N_{c}\left(t\right)\xi\left(t\right),\label{eq:clearCharAct}
\end{equation}
where $r_{0}$ is the relative impact of chartists' trading activity
and $N_{c}\left(t\right)$ is the number of chartists. In contrast
fundamentalists' demand is conditioned on their knowledge about the
market fundamentals, which is quantified as the fundamental price
$P_{f}$ (which has a physical dimension of generic price units, $\mathrm{p.u.}$):
\begin{equation}
D_{f}\left(t\right)=N_{f}\left(t\right)\ln\frac{P_{f}}{P\left(t\right)}=\left[N-N_{c}\left(t\right)\right]\ln\frac{P_{f}}{P\left(t\right)},\label{eq:clearFundAct}
\end{equation}
where $N_{f}\left(t\right)$ is the number of fundamentalists and
$P\left(t\right)$ is the current price. Note that here we assume
that the fundamental price is fixed, which is not true for the real
markets. Nevertheless adding variability to the fundamental price,
e.g., by assuming that it follows Brownian motion, would not have
a significant impact on the statistical properties of the model.

We have also assumed that a market maker instantaneously clears the
market by setting the price to the Walras equilibrium price, which
is obtained in the following manner:
\begin{equation}
D_{f}\left(t\right)+D_{c}\left(t\right)=0\quad\Rightarrow\quad P_{\text{eq}}(t)=P_{f}\exp\left(r_{0}\cdot\frac{N_{c}\left(t\right)}{N-N_{c}\left(t\right)}\cdot\xi\left(t\right)\right).\label{eq:clearPrice}
\end{equation}
If $\xi\left(t\right)$ fluctuates significantly faster than $N_{c}\left(t\right)$,
then the absolute return
\begin{equation}
\left|r\left(t\right)\right|=\left|\ln\frac{P\left(t\right)}{P\left(t-T\right)}\right|\propto\frac{N_{c}\left(t\right)}{N-N_{c}\left(t\right)}=y.
\end{equation}
Through out our papers we have referred to $y$ as modulating return
as it describes longer-term fluctuations of the absolute return, while
$\xi\left(t\right)$ dictates the rapid fluctuations and changes in
the sign of the return. In some of the earlier works $\xi\left(t\right)$
was even modeled as a noise \cite{Alfarano2005CompEco}.

Previously \cite{Kononovicius2012PhysA} we have also extended the
original herd behavior model by introducing the feedback of the modulating
return $y$ on the switching dynamics:
\begin{align}
p\left(N_{c}\rightarrow N_{c}+1\right) & =\left(N-N_{c}\right)\left[\sigma_{fc}+hN_{c}\right]\frac{\Delta t}{\tau\left(N_{c}\right)},\\
p\left(N_{c}\rightarrow N_{c}-1\right) & =N_{c}\left[\sigma_{cf}+h\left(N-N_{c}\right)\right]\frac{\Delta t}{\tau\left(N_{c}\right)},
\end{align}
where 
\begin{equation}
\tau\left(N_{c}\right)=\left(\frac{N_{c}}{N-N_{c}}\right)^{-\alpha}\equiv y^{-\alpha}.
\end{equation}
That is, $\tau\left(N_{c}\right)$ adjusts the characteristic time
scale of microscopic switching events according to the current global
value of the modulating return. Such feedback scenario implements
the coupling between returns and trading activity, which is well established
empirical fact \cite{Rak2013APP}. Note that in \cite{Kononovicius2012PhysA}
switching dynamics were assumed to correlate with trading activity.

Introduction of the feedback scenario enables us to obtain a more
general form of the SDE for $y$, which has tunable noise multiplicativity
exponent:
\begin{align}
\rmd y & =h\left[\varepsilon_{fc}+\left(2-\varepsilon_{cf}\right)y\right]\frac{1+y}{\tau\left(y\right)}\rmd t+\sqrt{\frac{2hy}{\tau\left(y\right)}}\left(1+y\right)\rmd W=\nonumber \\
 & =h\left[\varepsilon_{fc}+\left(2-\varepsilon_{cf}\right)y\right]\left(1+y\right)y^{\alpha}\rmd t+\sqrt{2hy^{1+\alpha}}\left(1+y\right)\rmd W\approx\\
 & \approx h\left(2-\varepsilon_{cf}\right)y^{2+\alpha}\rmd t+\sqrt{2hy^{3+\alpha}}\rmd W.\nonumber 
\end{align}

This SDE, assuming $y\gg1$, belongs to a class of SDEs exhibiting
power law statistics described in \cite{Ruseckas2010PhysRevE}. Thus
it the $y$ time series should exhibit power law statistics \cite{Kononovicius2012PhysA}:
\begin{equation}
P\left(y\right)\sim y^{-\varepsilon_{cf}-\alpha-1},\qquad S\left(f\right)\sim f^{-1-\frac{\varepsilon_{cf}+\alpha-2}{1+\alpha}}.\label{eq:ystats}
\end{equation}

This simple model already reproduces two main stylized facts. In the
later papers, e.g., \cite{Gontis2014PlosOne}, we have extended this
model by describing the mood dynamics using the same herd behavior
model. Though in order to fit the empirical absolute return PDF and
PSD we had to introduce exogenous noise, which we assumed to represent
additional randomness arising from the order book dynamics and possibly
an exogenous information inflow. In the next section we build the
order book model to address this assumption.

\subsection{Order book model with herd behavior}

Most of the ABMs, which consider statistical properties of the various
financial time series, directly or indirectly assume presence of the
market maker \cite{Cristelli2012Fermi}. While the real financial
markets are not cleared by an idealized market maker, most of the
contemporary financial markets implement trading by using the order
books. Similarly to the market makers order books record and execute
orders that the traders submit. The difference is that the orders
in the order book are executed only if there is an overlap between
the buy (bid) and the ask (sell) sides of the order book or if a market
order is submitted. While there is a significant body of literature
considering order book modeling \cite{Maslov2000PhysA,Challet2001PhysA,Preis2006EPL,Cristelli2010EPJB,Abergel2013IJTAF,Biondo2016JNTF,Donier2016JStatMech,Kanazawa2018PRL,Kanazawa2018PRE},
most of these models consider reproducing patterns observed at the
order book level and usually neglect stylized facts related to the
financial time series. It is worth to note that there are numerous
papers in economics, which suggest that the prices in various auctions
converge towards Walras equilibrium and that this convergence might
be comparatively fast \cite{Smith1962JPolEco,Bayati2015JEcoT,Nax2015IntJGameTh,Leshno2017SSRN,Kimbrough2018JEcoBehavOrg,Nax2018SSRN}.
Yet this convergence is not instantaneous and we might observe some
interesting effects in the high--frequency financial time series.
Here, while building our order book model, we will partly rely on
an empirically motivated order book model proposed by Kanazawa \textit{et
al.} \cite{Kanazawa2018PRL,Kanazawa2018PRE}.

As in \cite{Kanazawa2018PRL,Kanazawa2018PRE} we assume that chartists
as high--frequency traders submit unit volume limit orders to the
both sides of the order book. The submitted quotes, $Q_{i}^{\text{ask}}$
and $Q_{i}^{\text{bid}}$, are placed by the $i$-th agent the same
distance, $S_{i}$, away from the current valuation of the stock,
$V_{i}$:
\begin{align}
Q_{i}^{\text{ask}}\left(t\right) & =V_{i}\left(t\right)+S_{i},\\
Q_{i}^{\text{bid}}\left(t\right) & =V_{i}\left(t\right)-S_{i},\\
S_{i} & \sim\text{Gamma}\left(k,\theta\right),
\end{align}
where $k$ is the shape parameter of the Gamma distribution and $\theta$
is the scale parameter of the Gamma distribution. Here physical dimension
of $\theta$ is $\mathrm{p.u.}$ (generic price units) as it is also
the dimension of $Q_{i}$, $V_{i}$ and $S_{i}$. Further in this
paper we will use the empirically determined values $k=4$ and $\theta=15.5\,\mathrm{p.u.}$
(see \cite{Kanazawa2018PRL,Kanazawa2018PRE}) unless specified otherwise.
Note that these values were obtained specifically for USD/JPY exchange
rate in Forex and they might take different values for the other markets
or exchange rates. Yet we will rely on these values as best available
estimate at this point.

Next we will replace the sophisticated trend following mechanism originally
present in \cite{Kanazawa2018PRL,Kanazawa2018PRE} with a simpler
market order submission mechanism, which is similar to the one used
in the herd behavior model with instantaneous clearing. This simplification
leads to another simplifying assumption that valuations are homogeneous
$V_{i}\left(t\right)=V\left(t\right)$. Hence the valuation changes
only after a market order is executed. After a market order is executed,
the valuation is set to the value of realized quote. As in the original
order book approach, after the valuation is reset chartists update
their submitted quotes.

\begin{figure}
\begin{centering}
\includegraphics[width=0.7\textwidth]{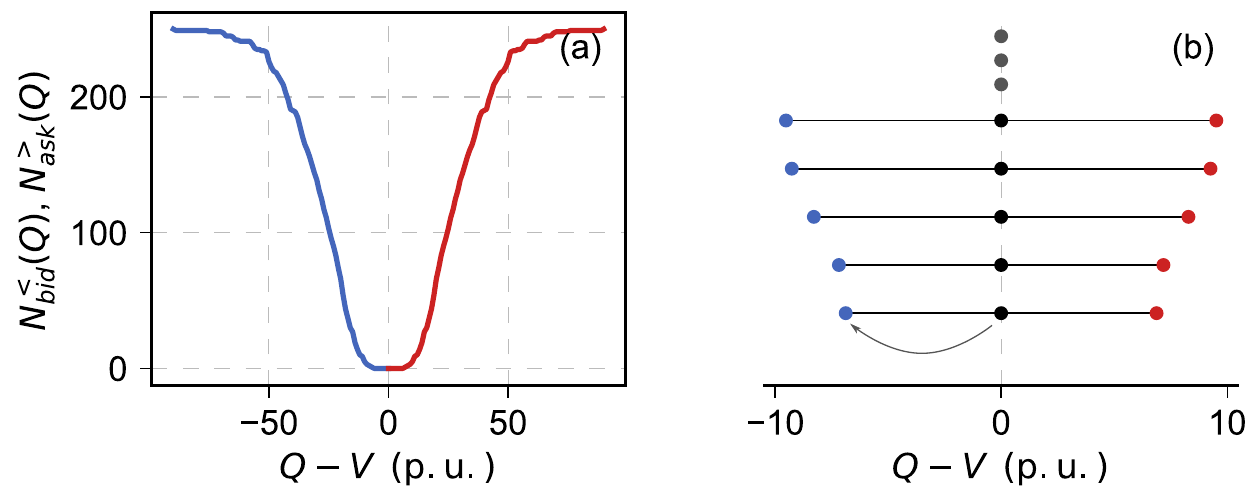}
\par\end{centering}
\caption{(color online) Full order book profile (a) and a zoom in to $5$ best
quotes present in the order book (b). Blue curve (a) and circles (b)
denote bid (buy) limit orders, red curve (a) and circles (b) denote
ask (sell) limit orders, while black circles (b) show $V_{i}$ for
each of the best quoters. In sub figure (a) ask profile is cumulative,
giving number of limit orders willing to sell for a given price or
larger, while bid profile is inversely cumulative, giving the number
of limit orders willing to buy at a given price or smaller. In subfigure
(b) the arrow shows where the valuation would move if market ask order
would be submitted.\label{fig:ob-profile}}
\end{figure}

An example of the full order book profile is shown in Fig.~\ref{fig:ob-profile}
(a). Sub figure (b) of Fig.~\ref{fig:ob-profile} provides zoomed
in picture of the $5$ best quotes, and shows the movement of valuation
if market ask order would be submitted. Blue and red circles represent
bid and ask quotes respectively, these circles are connected to show
that the both quotes are submitted by the same agent, whose valuation
is represented by the black circle. Note that under our simplifications
the profile is symmetric around quote equal to the current valuation,
which is identical for all agents. This would not be the case for
the original approach in \cite{Kanazawa2018PRL,Kanazawa2018PRE}.

Now let discuss the replacement of the trend following mechanism of
\cite{Kanazawa2018PRL,Kanazawa2018PRE} with a simpler one used in
the herd behavior model with instantaneous clearing. Let the chartists
submit unit volume market orders at rate
\begin{equation}
\lambda_{tC}\left(t\right)=\frac{\lambda_{e}}{\tau\left(N_{c}\left(t\right)\right)}\lambda_{tc}N_{c}\left(t\right),\label{eq:charAct}
\end{equation}
where $\lambda_{e}$ is the reference event rate (which has a physical
dimension of $\mathrm{s}^{-1}$) and $\lambda_{tc}$ -- the relative
market order submission rate by a single chartist agent. The submitted
market order is bid order with probability:
\begin{equation}
p_{\text{bid}}\left(t\right)=\frac{1+\xi\left(t\right)}{2},
\end{equation}
and ask market order is submitted otherwise, $p_{\text{ask}}\left(t\right)=1-p_{\text{bid}}\left(t\right)$.
Let us assume that the mood simply flips its sign at rate
\begin{equation}
\lambda_{M}\left(t\right)=\frac{\lambda_{e}}{\tau\left(N_{c}\left(t\right)\right)}\lambda_{m},
\end{equation}
where $\lambda_{m}$ is relative mood flipping rate. As the sign flip
does not change the modulus the mood will take only two possible values,
$\xi\left(t\right)\in\left\{ -\xi_{0},+\xi_{0}\right\} $ (here $\xi_{0}$
is initial value of the mood).

Fundamentalists are not present in \cite{Kanazawa2018PRL,Kanazawa2018PRE},
so we can keep our earlier assumptions about their behavior. Namely,
we assume that fundamentalists are willing to buy stock if there is
such ask quote for which $Q_{j}^{\text{ask}}\left(t\right)<P_{f}$,
and are willing to sell stock if there is such bid quote for which
$Q_{j}^{\text{bid}}\left(t\right)>P_{f}$. They will submit unit volume
market orders, if there are suitable quotes in the order book, at
rate:
\begin{equation}
\lambda_{tF}\left(t\right)=\frac{\lambda_{e}}{\tau\left(N_{c}\left(t\right)\right)}\lambda_{tf}\left[N-N_{c}\left(t\right)\right]\left|\ln\left(\frac{P\left(t\right)}{P_{f}}\right)\right|,\label{eq:fundAct}
\end{equation}
where $\lambda_{tf}$ is the relative market order submission rate
by a single fundamentalist agent.

Note that in \cite{Kanazawa2018PRL,Kanazawa2018PRE} power law return
distribution is obtained only after taking into account the fact that
the number of high--frequency traders (chartists) changes over time.
If the number of high--frequency traders (chartists) would be fixed,
then returns would be exponentially distributed. This intuition was
confirmed empirically by splitting the time series into two hour periods.
For each of these periods returns were found to be approximately exponentially
distributed, though the different values of rates were found for the
different periods. The values of rates were found to be positively
correlated with the average number of high--frequency traders (chartists)
present during the same periods. Next let us include this variation
to the current model by including switching behavior present in the
herd behavior model with instantaneous clearing.

It is straightforward to determine event rates for the trading strategy
switching: the fundamentalist will switch to the chartist trading
strategy at rate,
\begin{equation}
\lambda_{fc}\left(t\right)=\frac{\lambda_{e}}{\tau\left(N_{c}\left(t\right)\right)}\left[N-N_{c}\left(t\right)\right]\left[\varepsilon_{fc}+N_{c}\left(t\right)\right],
\end{equation}
while the chartist will switch to the fundamentalist trading strategy
at rate,
\begin{equation}
\lambda_{cf}\left(t\right)=\frac{\lambda_{e}}{\tau\left(N_{c}\left(t\right)\right)}N_{c}\left(t\right)\left[\varepsilon_{cf}+\left\{ N-N_{c}\left(t\right)\right\} \right].
\end{equation}
Note that these transition rates do not include parameter $h$, which
is because it is equivalent to $\lambda_{e}$. As soon as chartist
becomes fundamentalist his limit orders are canceled, also if fundamentalist
becomes chartist, then he immediately submits his limit orders.

As the number of agents in this model will always be finite, the probability
of $\tau$$\left(N_{c}\right)=0$ (or alternatively $N_{c}\left(t\right)=N$)
and $\tau\left(N_{f}\right)=\infty$ (or alternatively $N_{c}\left(t\right)=0$)
will be non--zero, which would lead to ``over--heating'' or ``freezing''
of the strategy switching dynamics. To avoid these edge cases let
us redefine the feedback scenario as:
\begin{equation}
\tau^{-1}\left(N_{c}\left(t\right)\right)=\lambda_{0}+\begin{cases}
\text{if }N_{c}\left(t\right)=N & \left[2N_{c}\left(t\right)\right]^{\alpha}\\
\text{else} & \left[\frac{N_{c}\left(t\right)}{N-N_{c}\left(t\right)}\right]^{\alpha}
\end{cases},
\end{equation}
where $\lambda_{0}$ is the relative minimum switching rate. In the
above we have multiplied $N_{c}$ by $2$ when taking the $N_{c}=N$
edge case into account, because the previous increase in $y$, number
of chartists increasing from $N_{c}=N-2$ to $N_{c}=N-1$, is approximately
double given large $N$.

We use the Gillespie algorithm \cite{Gillespie1977JPC,Gillespie1992AP}
to implement the order book model. The main idea behind the Gillespie
algorithm is that we can sum all of the event rates to obtain the
total event rate:
\begin{equation}
\lambda^{T}=\lambda_{cf}+\lambda_{fc}+\lambda_{M}+\lambda_{tF}+\lambda_{tC},
\end{equation}
which enables us to generate random inter--event times, which are
distributed exponentially:
\begin{equation}
\Delta t_{i}\sim\mathrm{Exp}\left(\lambda^{T}\right).
\end{equation}
After each $\Delta t_{i}$ one of the possible events happens. The
probability for any of the possible events to happen is proportional
to its rate:

\begin{equation}
p_{cf}=\frac{\lambda_{cf}}{\lambda^{T}},\quad p_{fc}=\frac{\lambda_{fc}}{\lambda^{T}},\quad p_{M}=\frac{\lambda_{M}}{\lambda^{T}},\quad p_{tF}=\frac{\lambda_{tF}}{\lambda^{T}},\quad p_{tC}=\frac{\lambda_{tC}}{\lambda^{T}}.
\end{equation}
As these probabilities sum to $1$, one of the five possible events
is bound to happen: either randomly selected chartist switches to
fundamentalist trading strategy (with probability $p_{cf}$) or randomly
selected fundamentalist switches to chartist trading strategy (with
probability $p_{fc}$) or the mood flips its sign (with probability
$p_{M}$) or the randomly selected fundamentalist submits market order
(with probability $p_{tF}$) and randomly selected chartist submits
market order (with probability $p_{tC}$).

\begin{figure}
\begin{centering}
\includegraphics[width=0.7\textwidth]{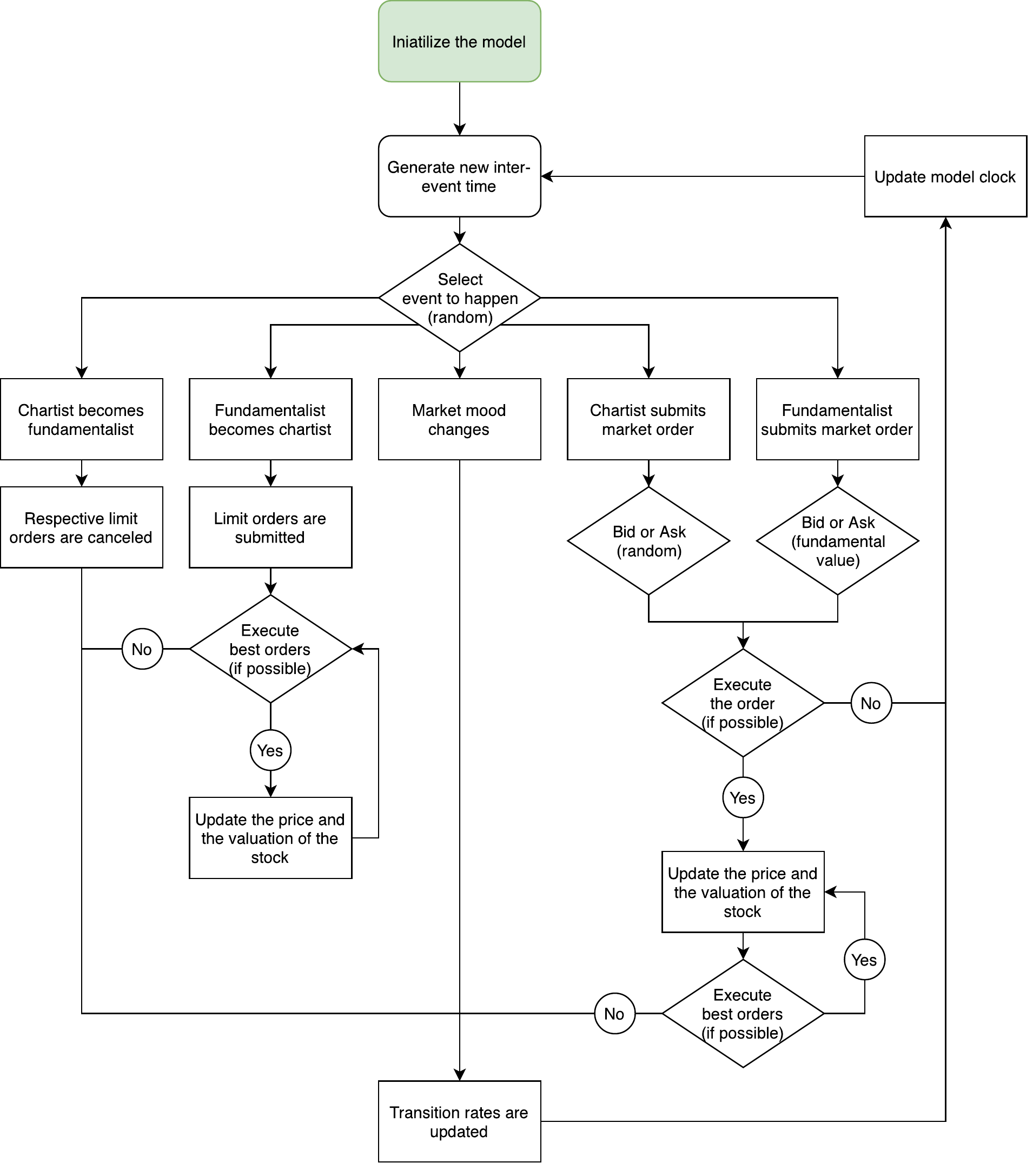}
\par\end{centering}
\caption{(color online) Flowchart illustrating the order book model with herd
behavior.\label{fig:schema}}

\end{figure}

The exact algorithm behind this model is summarized as a flowchart
in Fig.~\ref{fig:schema}. The code implementing this model is publicly
available on \url{https://github.com/akononovicius/herding-OB-model}.
All of the parameters used in this model are summarized in Table~\ref{tab:params}.

\begin{table}

\caption{List of parameters used in the order book model with herd behavior.\label{tab:params}}

\centering{}%
\begin{tabular}{|c|c|}
\hline 
\textbf{Symbol} & \textbf{Meaning}\tabularnewline
\hline 
\hline 
$N$ & number of agents\tabularnewline
\hline 
$\lambda_{e}$ & reference event rate (has physical dimension of $1/s$)\tabularnewline
\hline 
$\varepsilon_{cf}$ & relative idiosyncratic switching rate from chartists to fundamentalists\tabularnewline
\hline 
$\varepsilon_{fc}$ & relative idiosyncratic switching rate from fundamentalists to chartists\tabularnewline
\hline 
$\xi_{0}$ & absolute value of chartists' mood\tabularnewline
\hline 
$\lambda_{m}$ & relative rate at which mood flips its sign\tabularnewline
\hline 
$\lambda_{0}$ & relative minimum switching rate between chartists and fundamentalists\tabularnewline
\hline 
$\alpha$ & exponent of the feedback scenario\tabularnewline
\hline 
$k$ & order book shape parameter\tabularnewline
\hline 
$\theta$ & order book scale parameter (has physical dimension of generic price
units, $\mathrm{p.u.}$)\tabularnewline
\hline 
$P_{f}$ & fundamental price (has physical dimension of generic price units,
$\mathrm{p.u.}$)\tabularnewline
\hline 
$\lambda_{tc}$ & relative market order submission rate for chartists\tabularnewline
\hline 
$\lambda_{tf}$ & relative market order submission rate for fundamentalists\tabularnewline
\hline 
\end{tabular}
\end{table}

\subsection{Comparison between the models}

It is possible to approximately estimate equilibrium prices for the
order book model. From the discussion in the previous section we know
that chartists submit unit volume market orders at rate $\lambda_{tC}$,
with probability $p_{\text{bid}}$ they buy the stock and with probability
$1-p_{\text{bid}}$ they sell the stock. Hence their excess demand
(on average) is given by:
\begin{equation}
\bar{D}_{c}=\lambda_{tC}p_{\text{bid}}-\lambda_{tC}\left(1-p_{\text{bid}}\right)=\frac{\lambda_{e}}{\tau\left(N_{c}\left(t\right)\right)}\lambda_{tc}N_{c}\left(t\right)\xi\left(t\right).
\end{equation}
Note that the final result is similar to Eq.~(\ref{eq:clearCharAct}).
Fundamentalists on the other hand submit unit volume market orders
at rate $\lambda_{tF}$, they submit ask orders if there is such $j$
for which $Q_{j}^{\text{bid}}\left(t\right)>P_{f}$ and bid orders
if there is such $j$ for which $Q_{j}^{\text{ask}}\left(t\right)<P_{f}$.
Assuming that $P_{f}$ is not in the spread, the excess demand of
fundamentalists (on average) will be given by:
\begin{equation}
\bar{D}_{f}=\frac{\lambda_{e}}{\tau\left(N_{c}\left(t\right)\right)}\lambda_{tf}\left[N-N_{c}\left(t\right)\right]\ln\left(\frac{P\left(t\right)}{P_{f}}\right).
\end{equation}
Note that the final result is similar to Eq.~(\ref{eq:clearFundAct}).
Assuming that order book is non--empty and almost uniformly filled
we can obtain the equilibrium price:
\begin{equation}
\bar{D}_{c}+\bar{D}_{f}=0\quad\Rightarrow\quad P_{\text{eq}}(t)=P_{f}\exp\left(\frac{\lambda_{tc}}{\lambda_{tf}}\cdot\frac{N_{c}\left(t\right)}{N-N_{c}\left(t\right)}\cdot\xi\left(t\right)\right).
\end{equation}

\begin{figure}
\begin{centering}
\includegraphics[width=0.7\textwidth]{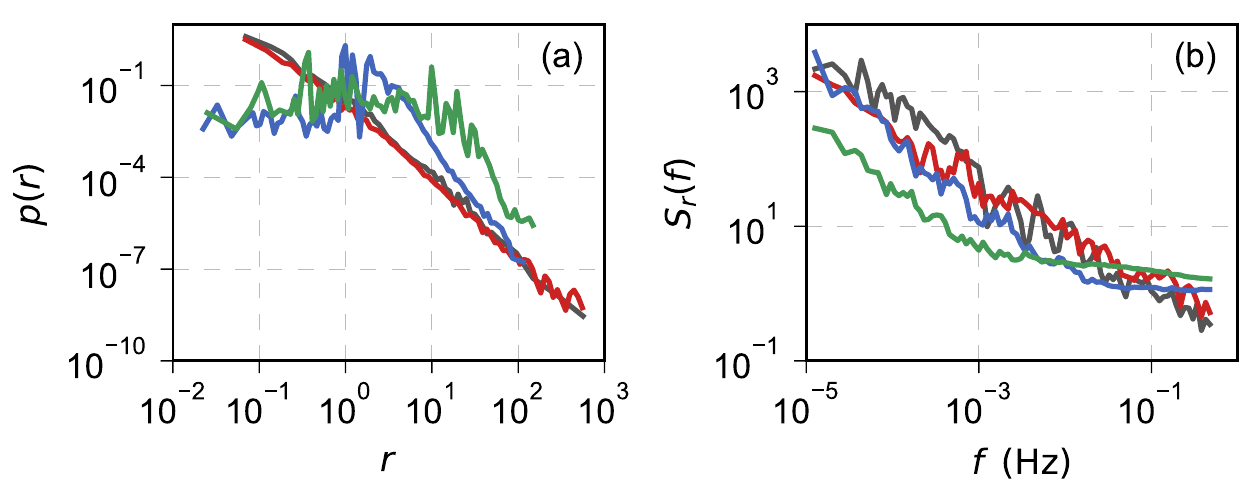}
\par\end{centering}
\caption{(color online) Comparison between the statistical properties, (a)
PDFs and (b) PSDs, of $y$ time series (gray curves) and absolute
return time series (red, blue and green curves), black curves show
the expected slopes of the statistical properties, as per Eq.~(\ref{eq:ystats}).
For the best comparison all of the time series were normalized to
unit standard deviation. The following parameter values were used
in numerical simulations: $N=500$, $\lambda_{e}=10^{-7}\,\mathrm{s}^{-1}$,
$\varepsilon_{cf}=\varepsilon_{fc}=1$, $\xi_{0}=0.2$, $\lambda_{m}=10^{3}$,
$\lambda_{0}=0.1$, $\alpha=1$, $k=4$, $\theta=15.5\,\mathrm{p.u.}$,
$P_{f}=3\cdot10^{4}\,\mathrm{p.u.}$ (all cases), $\lambda_{tc}=\lambda_{tf}=3\cdot10^{4}$
(red curves), $300$ (blue and gray curves), $3$ (green curves).\label{fig:largeLambda}}
\end{figure}

\begin{figure}
\begin{centering}
\includegraphics[width=0.7\textwidth]{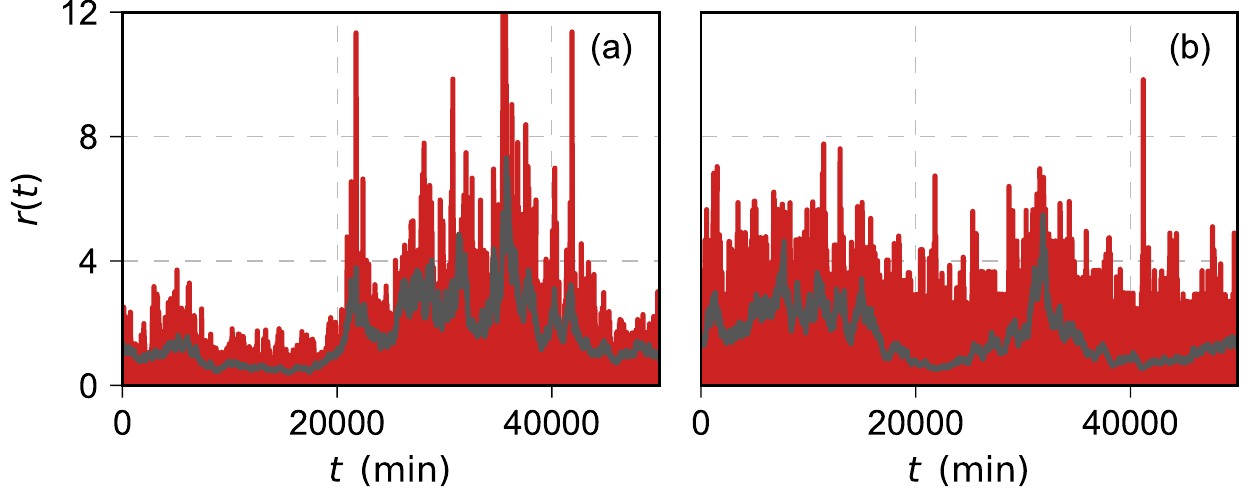}
\par\end{centering}
\caption{(color online) Sample fragments of the absolute return (red curves)
and $y$ (gray curves) time series. Parameter values are identical
to the ones used in Fig.~\ref{fig:largeLambda} except: (a) $\lambda_{tc}=\lambda_{tf}=3\cdot10^{4}$
(red curve in Fig.~\ref{fig:largeLambda}), (b) $\lambda_{tc}=\lambda_{tf}=3$
(green curve in Fig.~\ref{fig:largeLambda}). The correlation coefficients
between the samples are $\rho=0.67$ (a) and $0.03$ (b).\label{fig:returnFollow}}
\end{figure}

As the expression for the equilibrium price has the same form as in
Eq.~(\ref{eq:clearPrice}), we can expect that $y$ and return will
have similar statistical properties. The main condition we have to
ensure for the similarity to be observable is that enough trades happen
between $N_{c}$ changes, so that the equilibrium price could be reached.
This means that $\lambda_{tc}$ and $\lambda_{tf}$ have to be rather
large. As you can see in Fig.~\ref{fig:largeLambda} the agreement
between the statistical properties of $y$ and absolute return improves
as $\lambda_{tc}$ and $\lambda_{tf}$ grow larger. In Fig.~\ref{fig:returnFollow}
we have shown sample $y$ and absolute return time series, which can
be seen to correlate in sub figure (a) and be almost uncorrelated
in sub figure (b).

Correlation between $y$ and absolute return time series, see sample
series in Fig.~\ref{fig:returnFollow}, also is stronger with larger
$\lambda_{tc}$ and $\lambda_{tf}$. For the series shown in Fig.~\ref{fig:returnFollow}~(a)
correlation is strong $\rho\approx0.7$ (with $\lambda_{tc}=\lambda_{tf}=3\cdot10^{4}$),
while in Fig.~\ref{fig:returnFollow}~(a) correlation is negligible
$\rho\approx0$ (with $\lambda_{tc}=\lambda_{tf}=3$).

\begin{figure}
\begin{centering}
\includegraphics[width=0.7\textwidth]{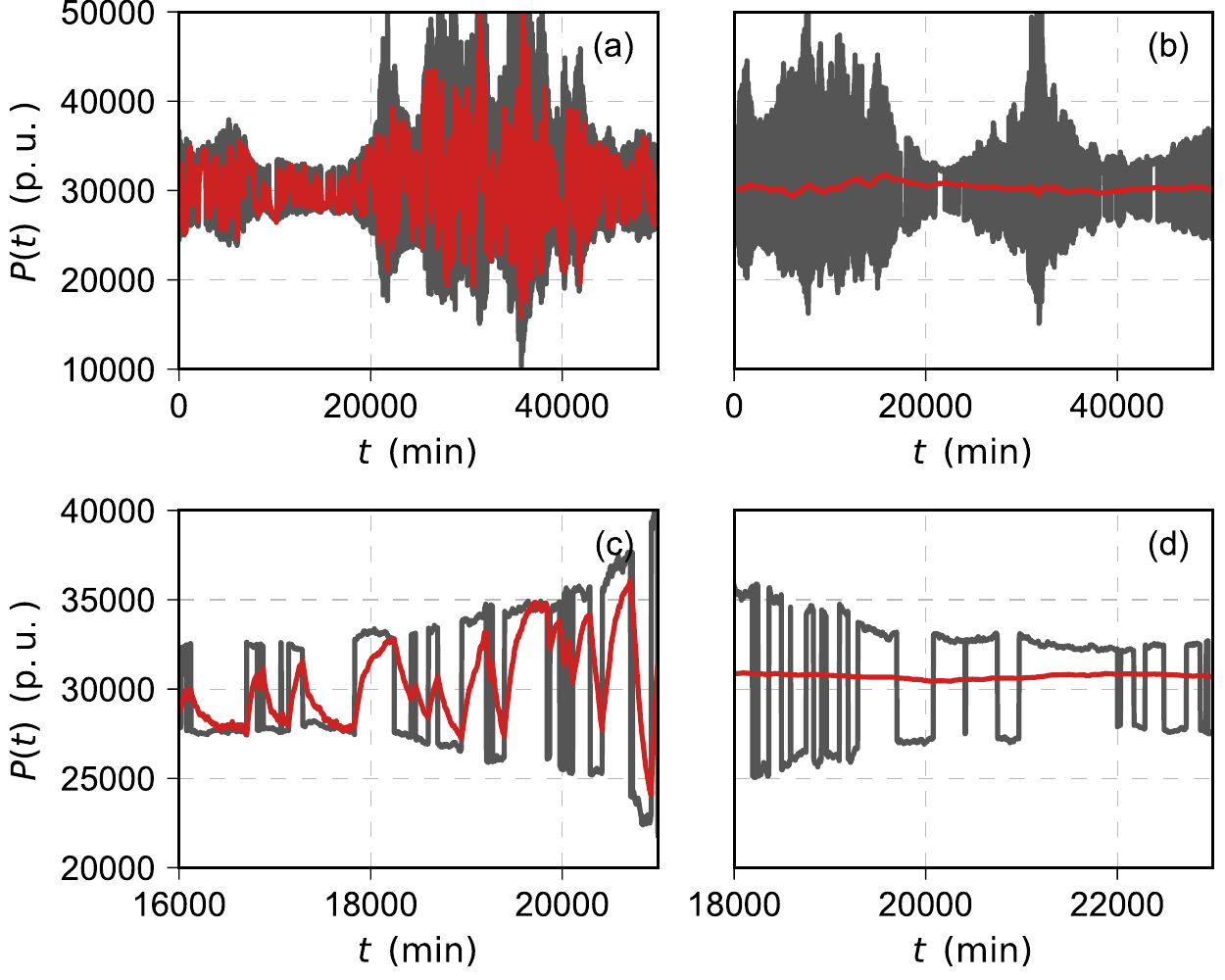}
\par\end{centering}
\caption{(color online) Sample fragments of the price (red curves) and the
equilibrium price (gray curves) time series. Parameter values are
identical to the ones used in Fig.~\ref{fig:largeLambda} except:
(a) and (c) $\lambda_{tc}=\lambda_{tf}=3\cdot10^{4}$ (red curve in
Fig.~\ref{fig:largeLambda}), (b) and (d) $\lambda_{tc}=\lambda_{tf}=3$
(green curve in Fig.~\ref{fig:largeLambda}).The correlation coefficients
between the samples are $\rho=0.54$ (a) and $0.02$ (b).\label{fig:priceFollow}}
\end{figure}

Similar intuition can be obtained from Fig.~\ref{fig:priceFollow}.
As you can see in (a) and (c), for large $\lambda_{tc}$ and $\lambda_{tf}$
(parameters the same as for the red curve in Fig.~\ref{fig:largeLambda})
the price tends to catch up with the changes in the equilibrium price.
Though the following is far from being perfect as can be seen by zooming
in on the series, (c). The correlation between the price and the equilibrium
price time series is mild. While for small $\lambda_{tc}$ and $\lambda_{tf}$,
(b) and (d), (parameters the same as for the green curve in Fig.~\ref{fig:largeLambda})
it is evident that the price does not manage to catch up with the
changes in the equilibrium price. As expected, there is almost no
correlation between the time series.

Based on these results we would like to argue that the fracture in
the PSD of the high--frequency absolute return time series happens
due to order book dynamics. Namely, the absolute return PSD in the
high frequency range is less sloped, because the markets are unable
to discover the equilibrium price that fast. It seems that it could
take a day or two (as the fracture is usually between $10^{-5}$ and
$10^{-4}$ Hz) for the markets to discover the new equilibrium price.

\section{Comparison against the empirical data\label{sec:empirical}}

Before making a comparison against the empirical data let us state
that neither order book model presented in \cite{Kanazawa2018PRL,Kanazawa2018PRE}
(which forms a basis of our order book approach) nor our financial
ABM \cite{Kononovicius2012PhysA,Gontis2014PlosOne} (which forms a
basis of our agent--based approach) is able to reproduce all of the
empirical statistical properties considered further in this Section.
Our earlier financial ABM \cite{Kononovicius2012PhysA,Gontis2014PlosOne}
was shown to reproduce statistical properties of absolute return reasonably
well, yet there was no way to consider trading activity in that model.
While \cite{Kanazawa2018PRL,Kanazawa2018PRE} has focused on reproduction
of order book dynamics only briefly touching upon statistical properties
of absolute return.

In this paper we use publicly available tick by tick trading data
from $12$ different Bitcoin exchanges. We have downloaded the data
from bitcoincharts.com website on July 5, 2018. List of the considered
Bitcoin time series is given in Table~\ref{tab:btc-data}. Note that
Coinbase and Kraken exchanges appear twice in the table, because they
contribute more than one exchange pair. These time series were selected,
because their data files were among top $5\%$ of the largest. Fisco's
BTC/JPY, Zaif's BTC/JPY and Zyado's BTC/EUR were also among top $5\%$
of the largest, but these time series were excluded, because their
statistical properties were too different from the rest of the time
series. Note that for the same reason we have truncated some of the
Bitcoin time series which remained under our consideration.

\begin{table}
\caption{List of the considered Bitcoin time series\label{tab:btc-data}}

\centering{}%
\begin{tabular*}{0.9\textwidth}{@{\extracolsep{\fill}}|c|c|c|c|r|}
\hline 
\textbf{exchange} & \textbf{exchange pair} & \textbf{period available} & \textbf{period used} & \textbf{trades}\tabularnewline
\hline 
\hline 
bitfinex & BTC/USD & 2013-03-31/2016-12-22 & from 2013-05-01 & $10121872$ ($99.4\%$)\tabularnewline
\hline 
bitflyer & BTC/JPY & 2017-07-04/2018-07-04 & whole & $30650659$\tabularnewline
\hline 
bitstamp & BTC/USD & 2011-09-13/2018-07-04 & except 2016-06-23 & $26065456$ ($99.9\%$)\tabularnewline
\hline 
btcbox & BTC/JPY & 2014-04-09/2018-07-04 & whole & $8900784$\tabularnewline
\hline 
btce & BTC/USD & 2011-08-14/2017-07-25 & whole & $32904793$\tabularnewline
\hline 
btcn & BTC/CNY & 2011-06-13/2017-09-30 & from 2013-04-01 & $114077056$ ($99.9\%$)\tabularnewline
\hline 
btcoid & BTC/IDR & 2014-02-09/2018-07-04 & whole & $8037858$\tabularnewline
\hline 
btctrade & BTC/CNY & 2013-05-19/2017-09-30 & whole & $19672943$\tabularnewline
\hline 
coinbase & BTC/EUR & 2015-04-23/2018-07-04 & whole & $15012224$\tabularnewline
\hline 
coinbase & BTC/USD & 2014-12-01/2018-07-04 & from 2017-05-01 & $30553110$ ($67.5\%$)\tabularnewline
\hline 
coincheck & BTC/JPY & 2014-10-31/2018-07-04 & except 2017-08-07 & $102124064$ ($99.9\%$)\tabularnewline
\hline 
kraken & BTC/EUR & 2014-01-08/2018-07-04 & whole & $21392263$\tabularnewline
\hline 
kraken & BTC/USD & 2014-01-07/2018-07-04 & whole & $10462050$\tabularnewline
\hline 
okcoin & BTC/CNY & 2013-06-12/2015-04-05 & whole & $99999546$\tabularnewline
\hline 
\end{tabular*}
\end{table}

\begin{figure}
\begin{centering}
\includegraphics[width=0.7\textwidth]{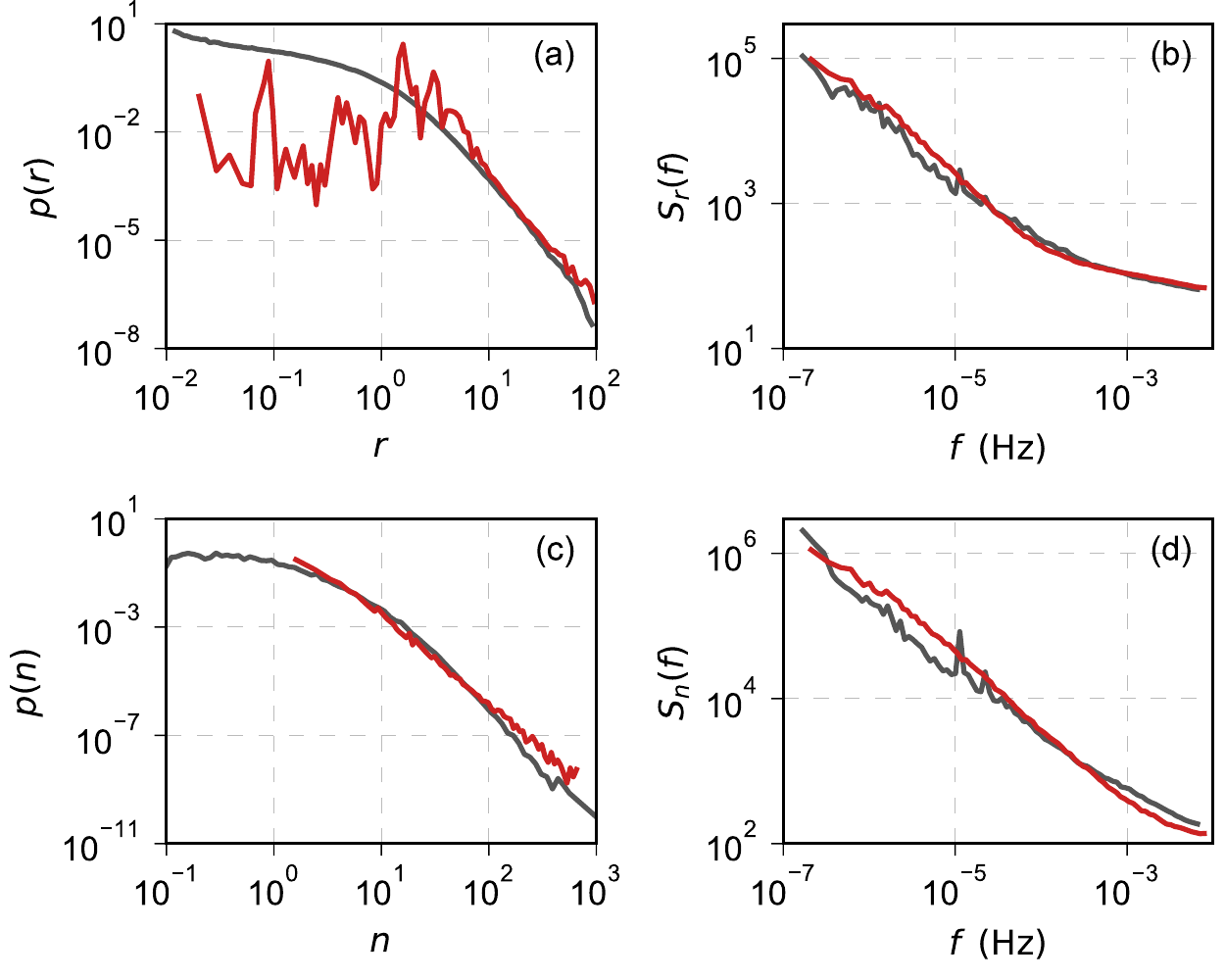}
\par\end{centering}
\caption{(color online) Comparison between the empirical Bitcoin statistical
properties (gray curves) and statistical properties generated by the
model (red curves): (a) one minute absolute return PDF, (b) one minute
absolute return PSD, (c) trading activity per one minute PDF, (d)
trading activity per one minute PSD. The following parameter set was
used in numerical simulations: $N=500$, $\lambda_{e}=10^{-7}\,\mathrm{s}^{-1}$,
$\varepsilon_{fc}=5$, $\varepsilon_{cf}=2$, $\xi_{0}=0.2$, $\lambda_{m}=10$,
$\lambda_{tc}=25$, $\lambda_{tf}=75$, $\lambda_{0}=0.4$, $\alpha=2$,
$k=4$, $\theta=15.5\,\mathrm{p.u.}$, $P_{f}=3\cdot10^{4}\,\mathrm{p.u.}$.\label{fig:model-btc}}
\end{figure}

For each of the considered Bitcoin time series we have produced one
minute absolute return (normalized to the standard deviation) and
trading activity (defined as trades per time interval and normalized
to the mean) time series. For each of the produced one minute time
series we have calculated PDF and PSD. The obtained statistical properties
were averaged to produce average profile for each of the statistical
properties. To select the model parameters we have used simulated
annealing technique with a goal to reproduce these averaged statistical
properties. As you can see in Fig.~\ref{fig:model-btc} the obtained
agreement is rather good.

Earlier works, such as the ones found in \cite{Drozdz2018Chaos,Begusic2018PhysA},
has already carried out detailed analysis of the Bitcoin time series
and established that the stylized facts for the Bitcoin are somewhat
different from the stylized facts established for the stocks. Namely,
\cite{Begusic2018PhysA} has reported that the Bitcoin returns exhibit
heavier tails than ordinary stock returns. Similar finding was also
reported in \cite{Drozdz2018Chaos}. Yet these works disagree on whether
the tail index changes over time. In \cite{Begusic2018PhysA} it was
reported that while the tail index has slightly increased, but not
enough statistical evidence was found to formally claim that the tail
index is increasing. In \cite{Drozdz2018Chaos}, on the other hand,
a significant change in the tail index is reported. Furthermore other
statistical properties of absolute return, such as slope of the auto--correlation
function, Hurst exponent and multi-scaling properties, seem to approach
values observed for the ordinary stocks. This is interpreted as a
sign that Bitcoin market is becoming ``mature'' market. In this paper
we did not carry out a rigorous empirical analysis, at least as rigorous
as in \cite{Drozdz2018Chaos,Begusic2018PhysA}. We have just checked
whether PDFs and PSDs of absolute return and trading activity change
over time and found that the change, if present, is negligible for
the exchange rates that have remained under our consideration. While
the return tail index seems to be similar to the one obtained for
NYSE stocks under our consideration ($\lambda\approx3.7$; analysis
of NYSE stocks are discussed in the next paragraph). A thorough analysis
of the Bitcoin time series or a meta analysis of the methods used
in \cite{Drozdz2018Chaos,Begusic2018PhysA} would be due, but this
topic is out of the scope of this paper.

\begin{figure}
\begin{centering}
\includegraphics[width=0.7\textwidth]{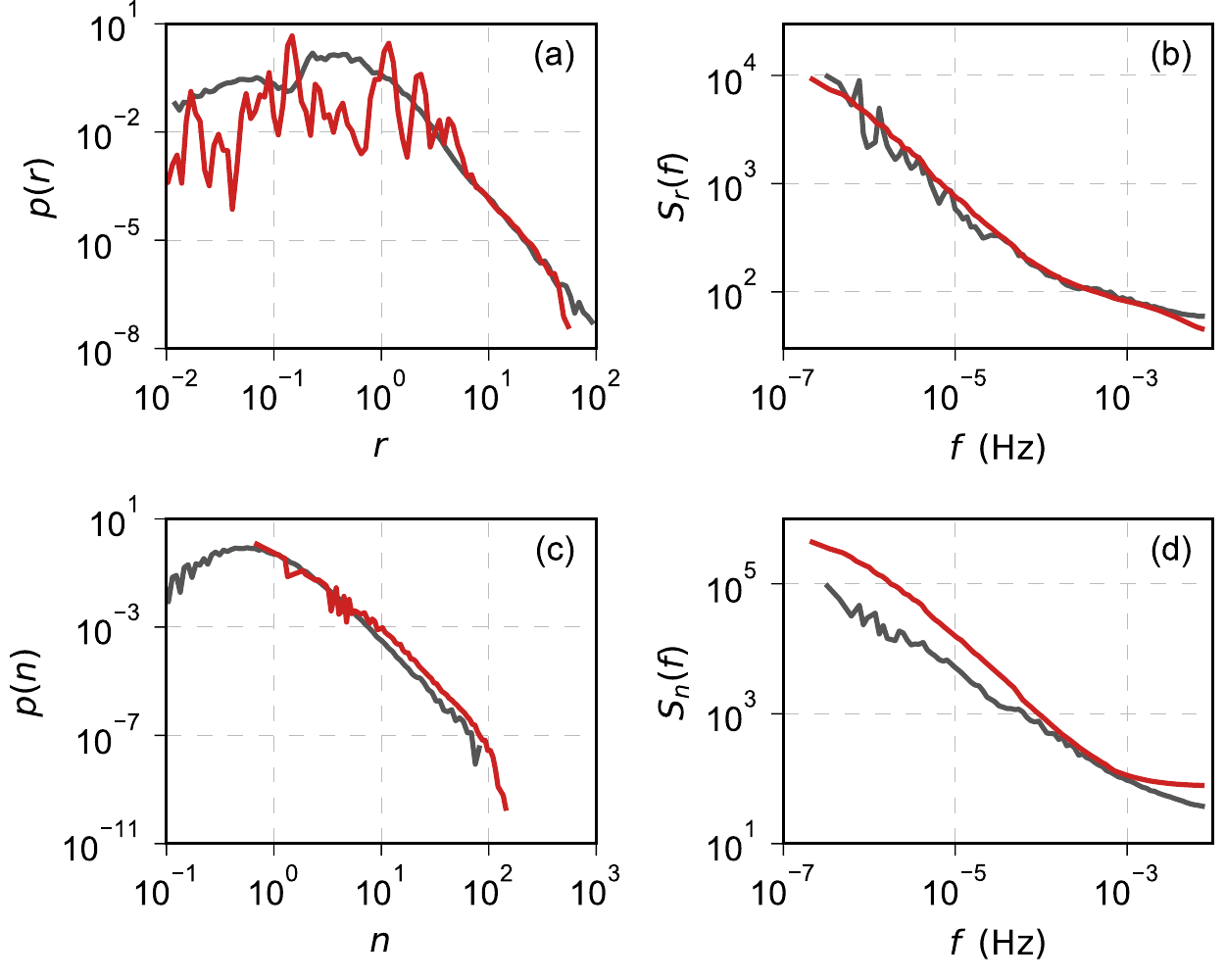}
\par\end{centering}
\caption{(color online) Comparison between the empirical NYSE stocks' statistical
properties (gray curves) and statistical properties generated by the
model (red curves). Parameter values are identical to the ones used
in Fig.~\ref{fig:model-btc} except: $\xi_{0}=1$, $\lambda_{tc}=2$,
$\lambda_{0}=1.5$.\label{fig:model-nyse}}
\end{figure}

We have also considered the statistical properties of 26 tickers from
NYSE. The considered tickers include: ABT, ADM, BA, BMY, C, CVX, DOW,
FON, FNM, GE, GM, HD, IBM, JNJ, JPM, KO, LLY, MMM, MO, MOT, MRK, SLE,
PFE, T, WMT and XOM. All their time frames are from January, 2005
to March, 2007. As with the Bitcoin time series, we have obtained
averaged statistical properties for NYSE data set. Using simulated
annealing technique we have obtained another best fit parameter set
for our model. To obtain a better fit we had to divide absolute return
time series generated by the model by factor of $3$. This indicates
that the model still lacks something, though it seems to reproduce
correct behavior for the tail of the distribution. As we can see in
Fig.~\ref{fig:model-nyse} after this correction the agreement between
the model and the data appears to be rather good. The obtained parameter
set is similar to the one obtained for the Bitcoin case. Though there
are some differences. The mood swings seem to be larger in NYSE case
(the respective $\xi_{0}$ is larger), while Bitcoin trading would
seem to be more random (which is somewhat consistent with assumptions
made in \cite{Cocco2017JEIC,Cocco2019FI}). On the other hand chartists
seem to submit less order in NYSE case ($\lambda_{tc}$ is smaller),
which would indicate prices in NYSE a more impacted by the market
fundamentals. Base trading activity seems to be higher, $\lambda_{0}$
is larger, for NYSE. The last two differences could potentially indicate
that NYSE is a more mature market than Bitcoin exchanges, as is also
noted based on empirical observations in \cite{Drozdz2018Chaos}.

\section{Impact of the model parameters\label{sec:params}}

In this section we check to see how changing the model's parameter
values impact the statistical properties of absolute return and trading
activity generated by the model. In all figures in this section we
will show three curves. Usually one will be generated with a larger
parameter value than used to produce Fig.~\ref{fig:model-btc} (blue
curve), one smaller (green curve) and one identical (red curve).

\begin{figure}
\begin{centering}
\includegraphics[width=0.7\textwidth]{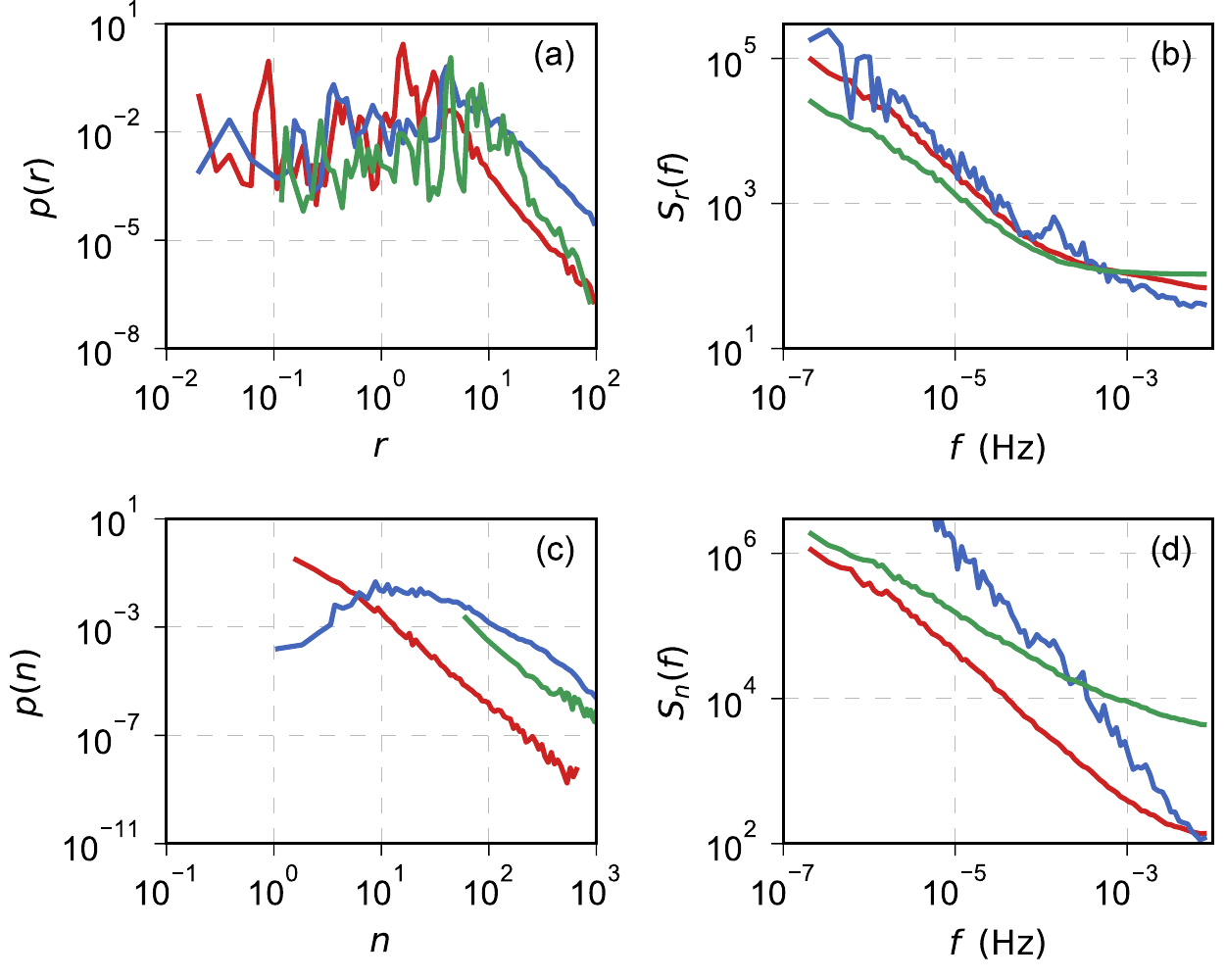}
\par\end{centering}
\caption{(color online) Influence of $N$ parameter on the model's statistical
properties. Parameter values are identical to the ones used in Fig.~\ref{fig:model-btc}
except: $N=5000$ (blue curve), $N=50$ (green curve).\label{fig:params-n}}
\end{figure}

Models built on the non--extensive formulation of the herd behavior
model are known to avoid the $N$--dependence problem \cite{Alfarano2005CompEco,Alfarano2008Dyncon,Alfarano2009Dyncon,Flache2011JCR,Kononovicius2012PhysA,Kononovicius2014EPJB,Gontis2014PlosOne,Peralta2018VMNet},
but as we can see in Fig.~\ref{fig:params-n} this model has some
kind of $N$--dependence. Yet the fluctuations do not disappear with
larger $N$, it seems that the model starts to exhibit even fatter
tails. This is most likely occurs due to the implemented mood mechanism:
the more agents and the more chartist agents, the more mood is reflected
in the time series.

\begin{figure}
\begin{centering}
\includegraphics[width=0.7\textwidth]{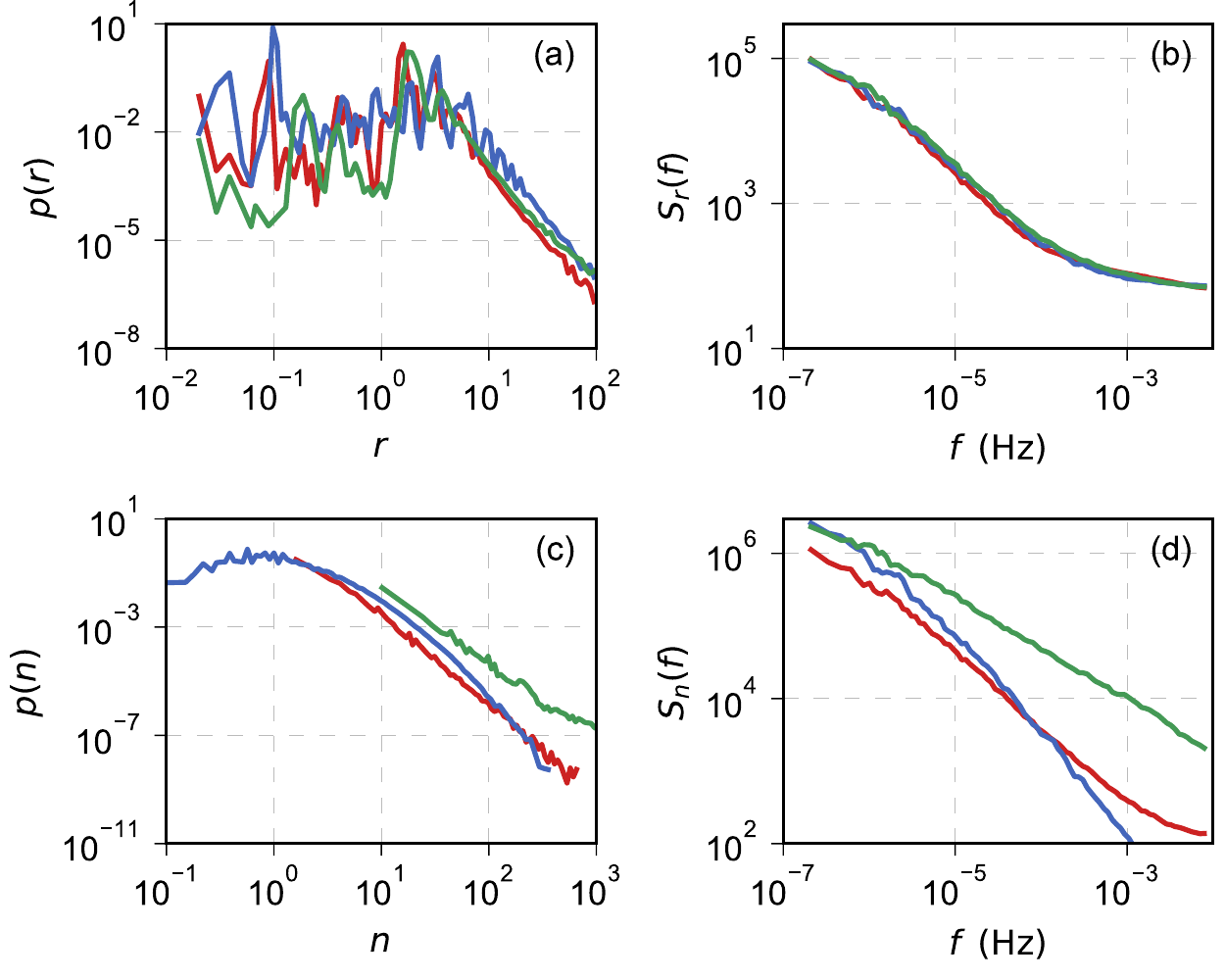}
\par\end{centering}
\caption{(color online) Influence of $N$ parameter on the model's statistical
properties, when $N\lambda_{tc}=\text{const}$ and $N\lambda_{tf}=\text{const}$.
Parameter values are identical to the ones used in Fig.~\ref{fig:model-btc}
except: $N=5000$, $\lambda_{tc}=2.5$, $\lambda_{tf}=7.5$ (blue
curve), $N=50$, $\lambda_{tc}=250$, $\lambda_{tf}=750$ (green curve).\label{fig:params-n-la}}
\end{figure}

We can partly eliminate this dependence be recalling that while $y$
dynamics are not influenced by $N$, but the number of trades per
time window does depend on $N$. By requiring that $N\lambda_{tc}=\text{const}$
and $N\lambda_{tf}=\text{const}$ we eliminate this dependence. And
as we can see in Fig.~\ref{fig:params-n-la} then changing $N$ does
not influence the statistical properties of absolute return. Though
$N$ retains the impact on the trading activity.

\begin{figure}
\begin{centering}
\includegraphics[width=0.7\textwidth]{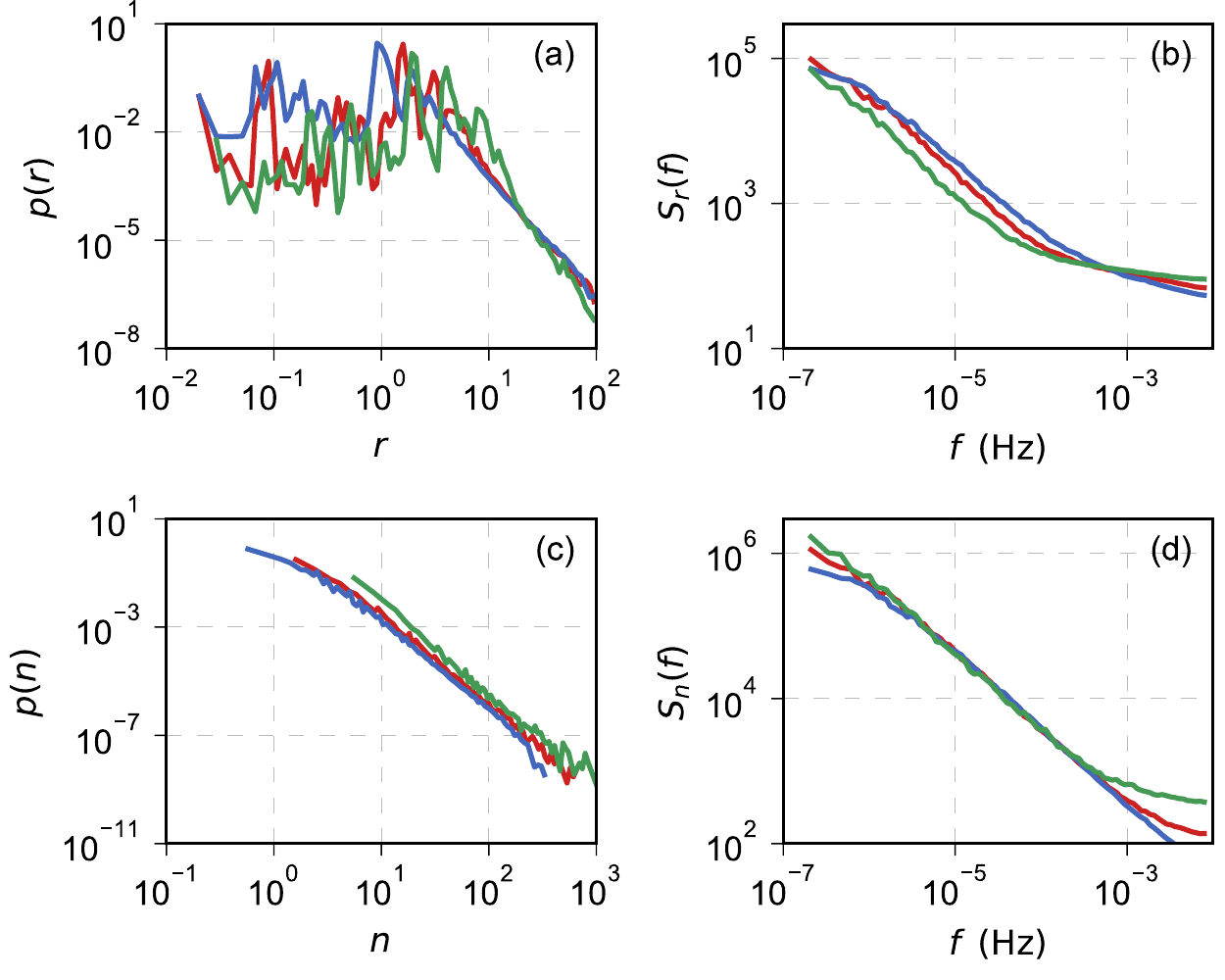}
\par\end{centering}
\caption{(color online) Influence of $\lambda_{e}$ parameter on the model's
statistical properties. Parameter values are identical to the ones
used in Fig.~\ref{fig:model-btc} except: $\lambda_{e}=3\cdot10^{-7}\,\mathrm{s}^{-1}$
(blue curve), $\lambda_{e}=3\cdot10^{-8}\,\mathrm{s}^{-1}$ (green
curve).\label{fig:params-lae}}
\end{figure}

Changing $\lambda_{e}$ parameter value seems to have a similar impact
as changing $h$ in the original herd behavior model (see Fig.~\ref{fig:params-lae}).
Namely the PSD of the absolute return shifts to the right as we increase
$\lambda_{e}$. Though due to the absolute return formula comparing
log--prices at two different points in time, the PDF of the absolute
return might also be impacted: it seem that the PDF might obtain heavier
tails as $\lambda_{e}$ becomes larger. Interestingly changing $\lambda_{e}$
does not seem to have any qualitative effect on the statistical properties
of the trading activity. Most likely increasing $\lambda_{e}$ simply
increase the mean of trades per time window without changing anything
else.

\begin{figure}
\begin{centering}
\includegraphics[width=0.7\textwidth]{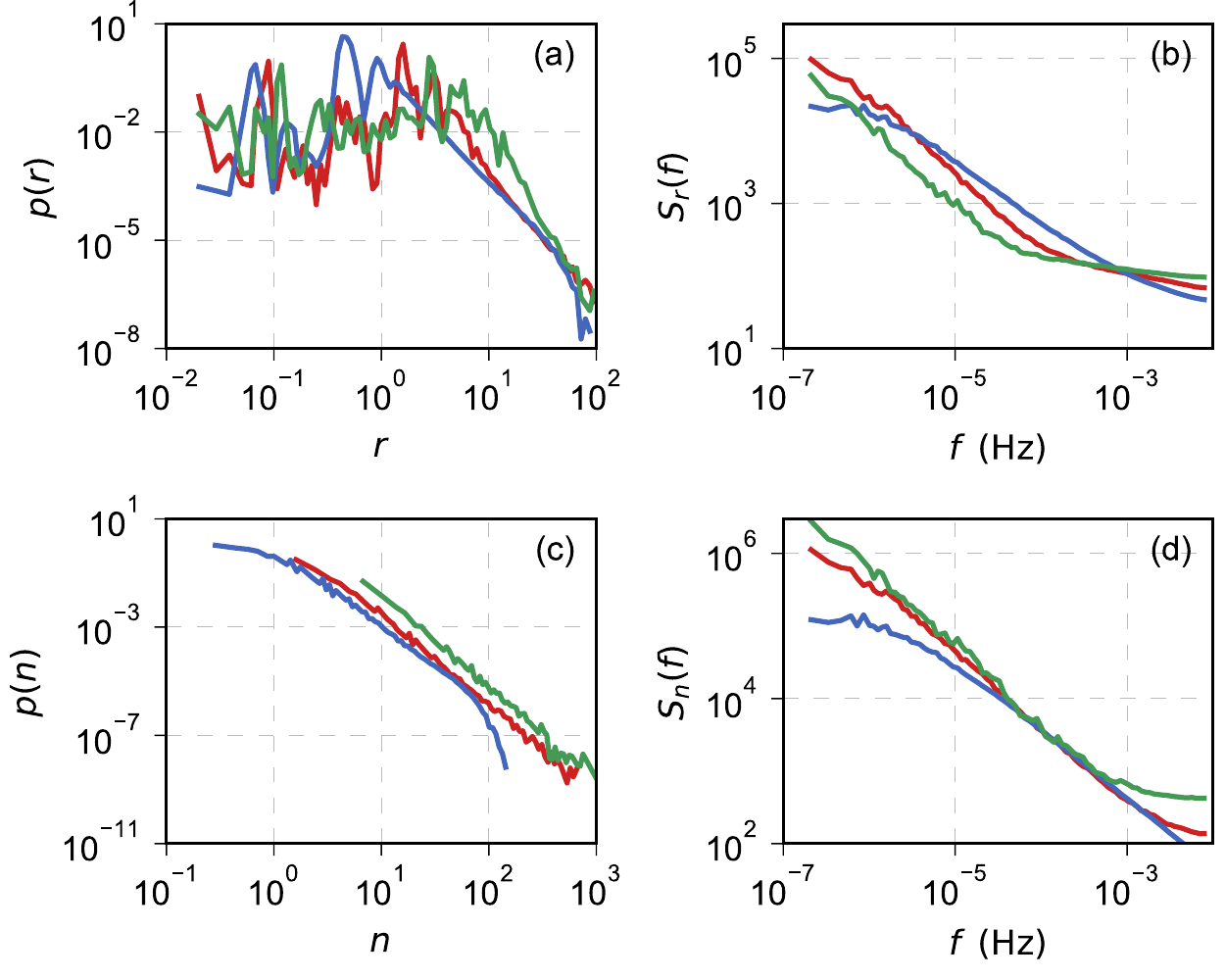}
\par\end{centering}
\caption{(color online) Influence of $\varepsilon_{fc}$ parameter on the model's
statistical properties. Parameter values are identical to the ones
used in Fig.~\ref{fig:model-btc} except: $\varepsilon_{fc}=15$
(blue curve), $\varepsilon_{fc}=0.7$ (green curve).\label{fig:params-efc}}
\end{figure}

Parameter $\varepsilon_{fc}$ does not have significant impact on
the statistical properties of the model with instantaneous clearing,
Eq.~\ref{eq:ystats}, but it seems that it is able to impact the
statistical properties of the absolute return in the order book model
(see Fig.~\ref{fig:params-efc}). As $\varepsilon_{fc}$ increases
the tails of the PDF becomes heavier and the PSD becomes flatter.
The statistical properties of the trading activity do not seem to
change qualitatively, the tail of the PDF remains the same nor does
the steepness of the PSD change. Most likely larger $\varepsilon_{fc}$
simply increases mean trading activity.

\begin{figure}
\begin{centering}
\includegraphics[width=0.7\textwidth]{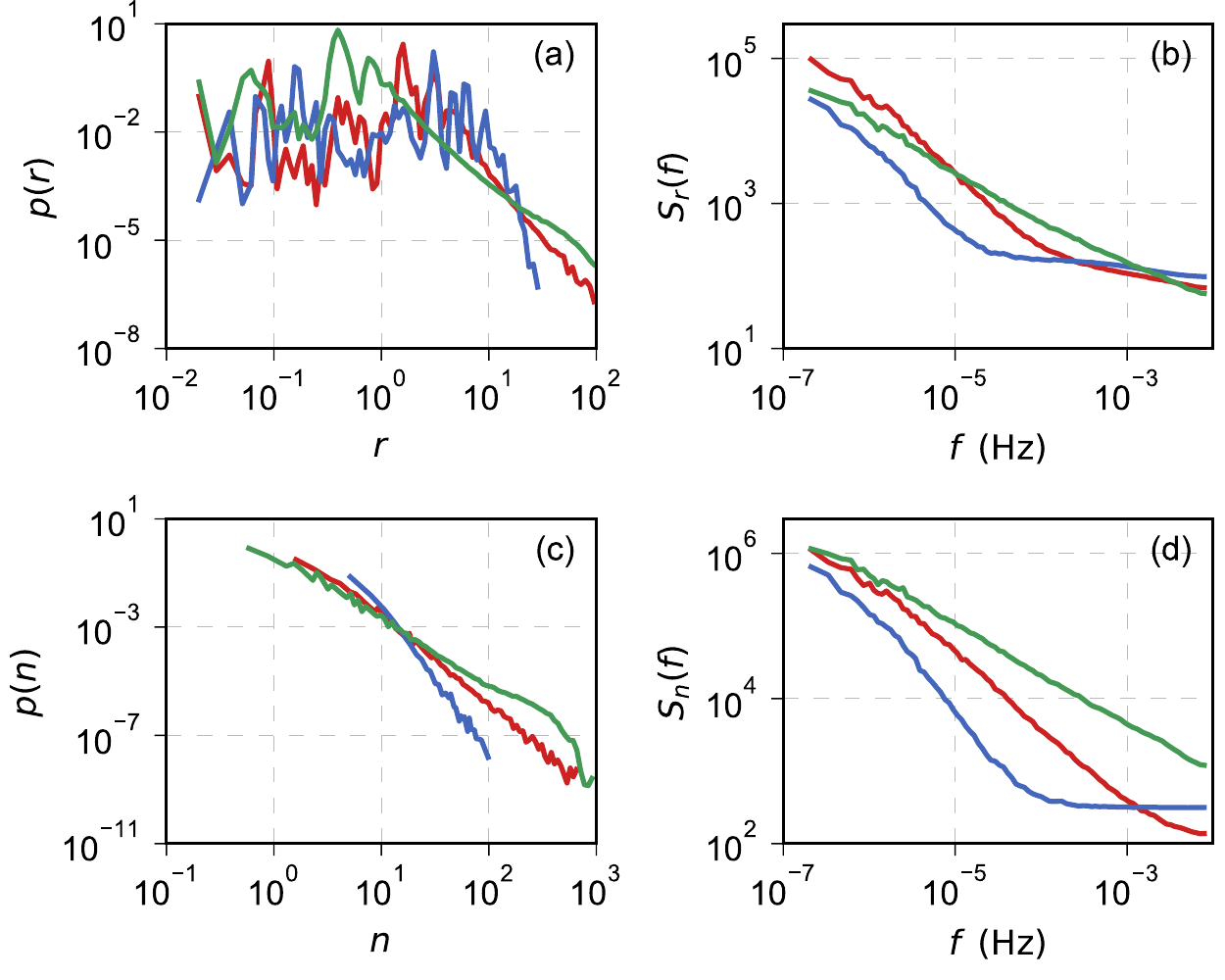}
\par\end{centering}
\caption{(color online) Influence of $\varepsilon_{cf}$ parameter on the model's
statistical properties. Parameter values are identical to the ones
used in Fig.~\ref{fig:model-btc} except: $\varepsilon_{cf}=6$ (blue
curve), $\varepsilon_{cf}=0.7$ (green curve).\label{fig:params-ecf}}
\end{figure}

Parameter $\varepsilon_{cf}$ seems to have the opposite effect on
the statistical properties of absolute return (see Fig.~\ref{fig:params-ecf}).
As $\varepsilon_{cf}$ increases the tail of the PDF become lighter,
while the slope of the PSD becomes steeper. The impact on the statistical
properties of the trading activity seems to be both quantitative,
the mean number of trades per time window decreases as $\varepsilon_{cf}$
increases, and qualitative, the tail of the PDF becomes lighter and
the slope of the PSD becomes steeper. These effects are most likely
caused by the dynamics reflected by the model with instantaneous clearing
as such dependence is predicted by Eq.~\ref{eq:ystats}.

\begin{figure}
\begin{centering}
\includegraphics[width=0.7\textwidth]{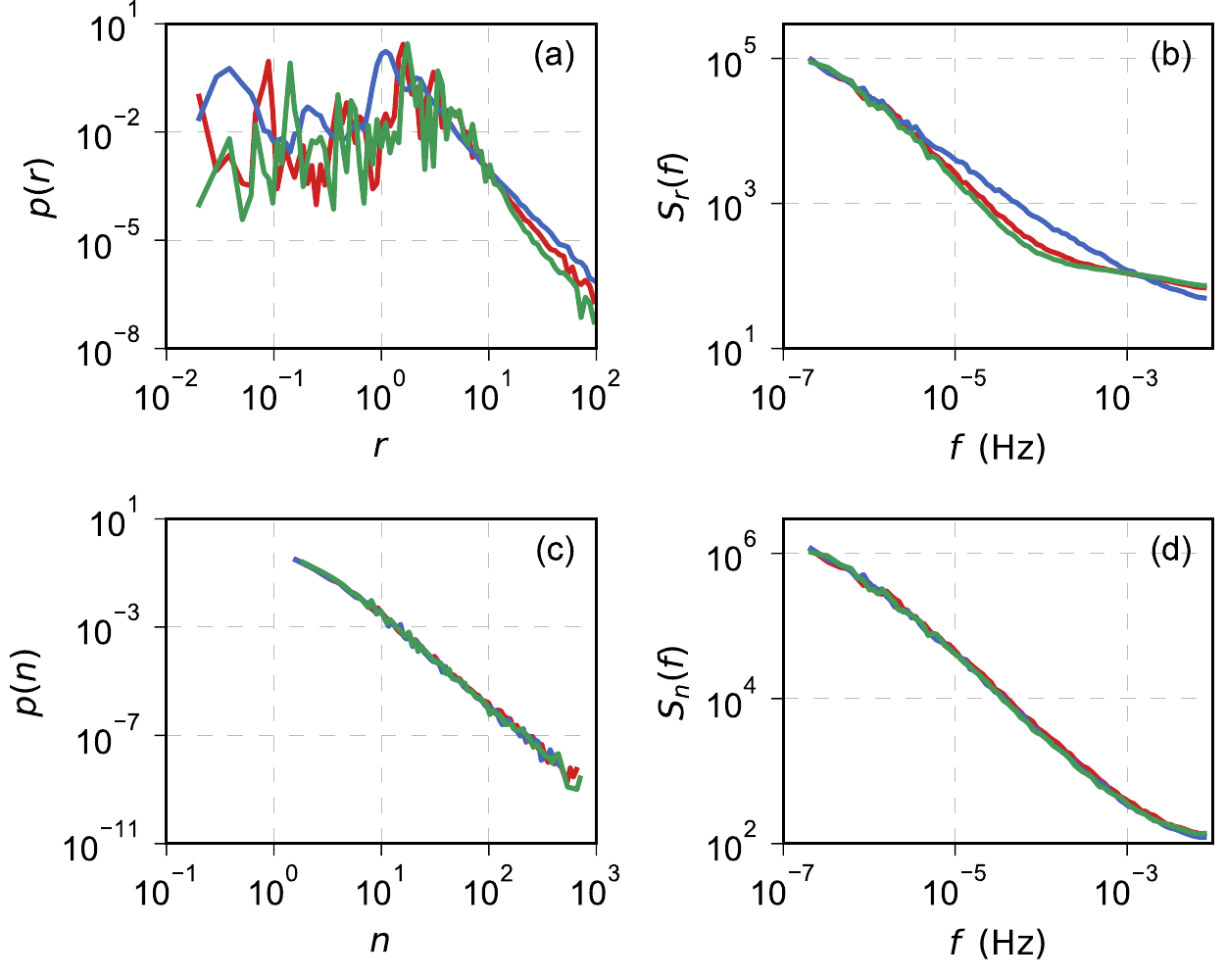}
\par\end{centering}
\caption{(color online) Influence of $\xi_{0}$ parameter on the model's statistical
properties. Parameter values are identical to the ones used in Fig.~\ref{fig:model-btc}
except: $\xi_{0}=0.6$ (blue curve), $\xi_{0}=0.1$ (green curve).\label{fig:params-xi}}
\end{figure}

\begin{figure}
\begin{centering}
\includegraphics[width=0.7\textwidth]{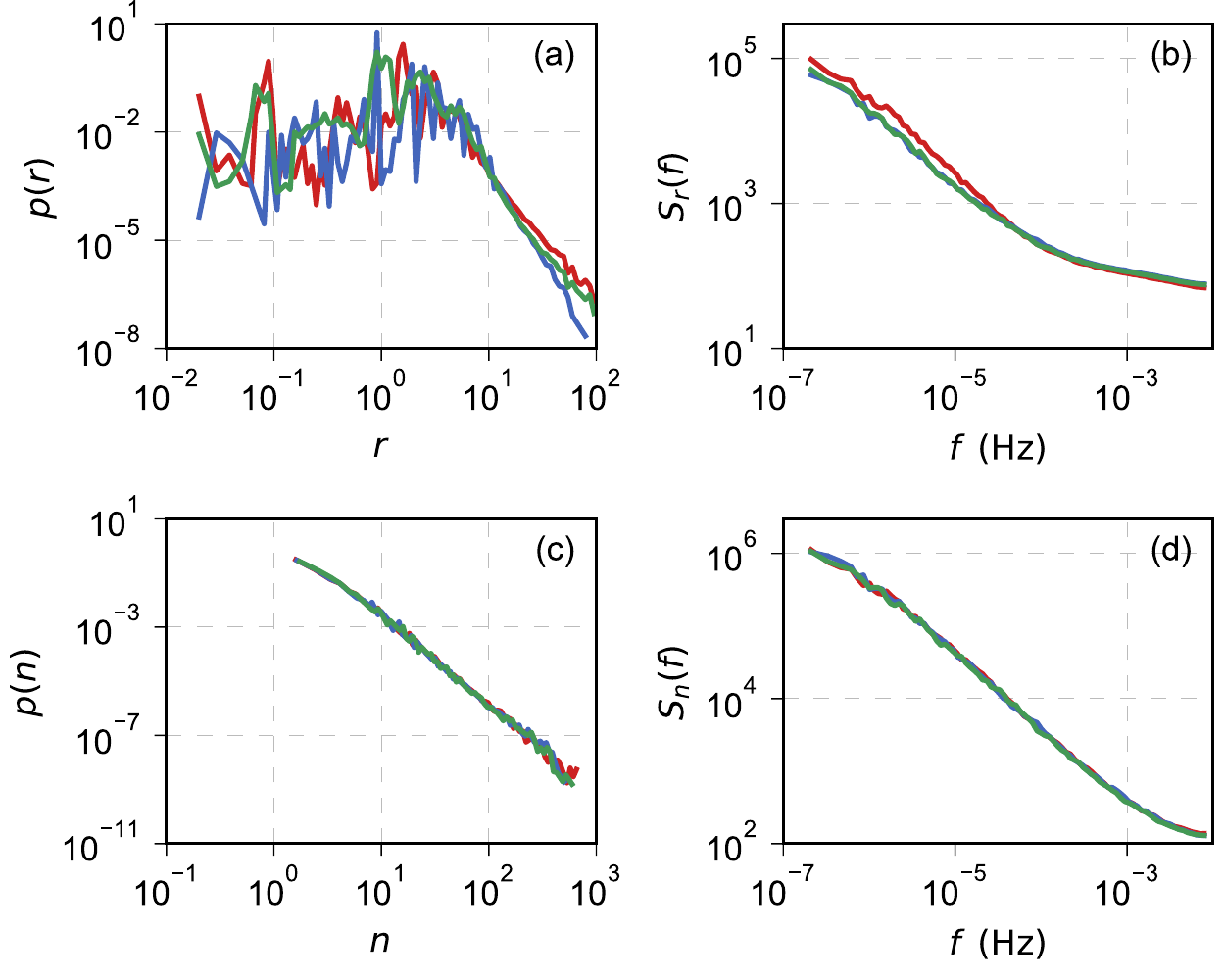}
\par\end{centering}
\caption{(color online) Influence of $\lambda_{m}$ parameter on the model's
statistical properties. Parameter values are identical to the ones
used in Fig.~\ref{fig:model-btc} except: $\lambda_{m}=300$ (blue
curve), $\lambda_{m}=0.3$ (green curve).\label{fig:params-lamc}}
\end{figure}

The mood dynamics, $\xi_{0}$ and $\lambda_{mc}$, doesn't seem to
have a significant impact on the statistical properties of both absolute
return and trading activity (see Figs.~\ref{fig:params-xi} and \ref{fig:params-lamc}).
Though small effect of $\xi_{0}$ on the absolute return PDF and PSD
are visible. Larger $\xi_{0}$ makes the tail of the PDF less fat
and the PSD flatter. Changing $\lambda_{mc}$ would have a larger
if the base value of $\xi_{0}$ was larger. Increasing $\lambda_{mc}$
would have similar effect as decreasing $\xi_{0}$ as with larger
$\lambda_{m}$ the effective mood (the average trend) would be smaller
than than the true value of $\xi_{0}$.

\begin{figure}
\begin{centering}
\includegraphics[width=0.7\textwidth]{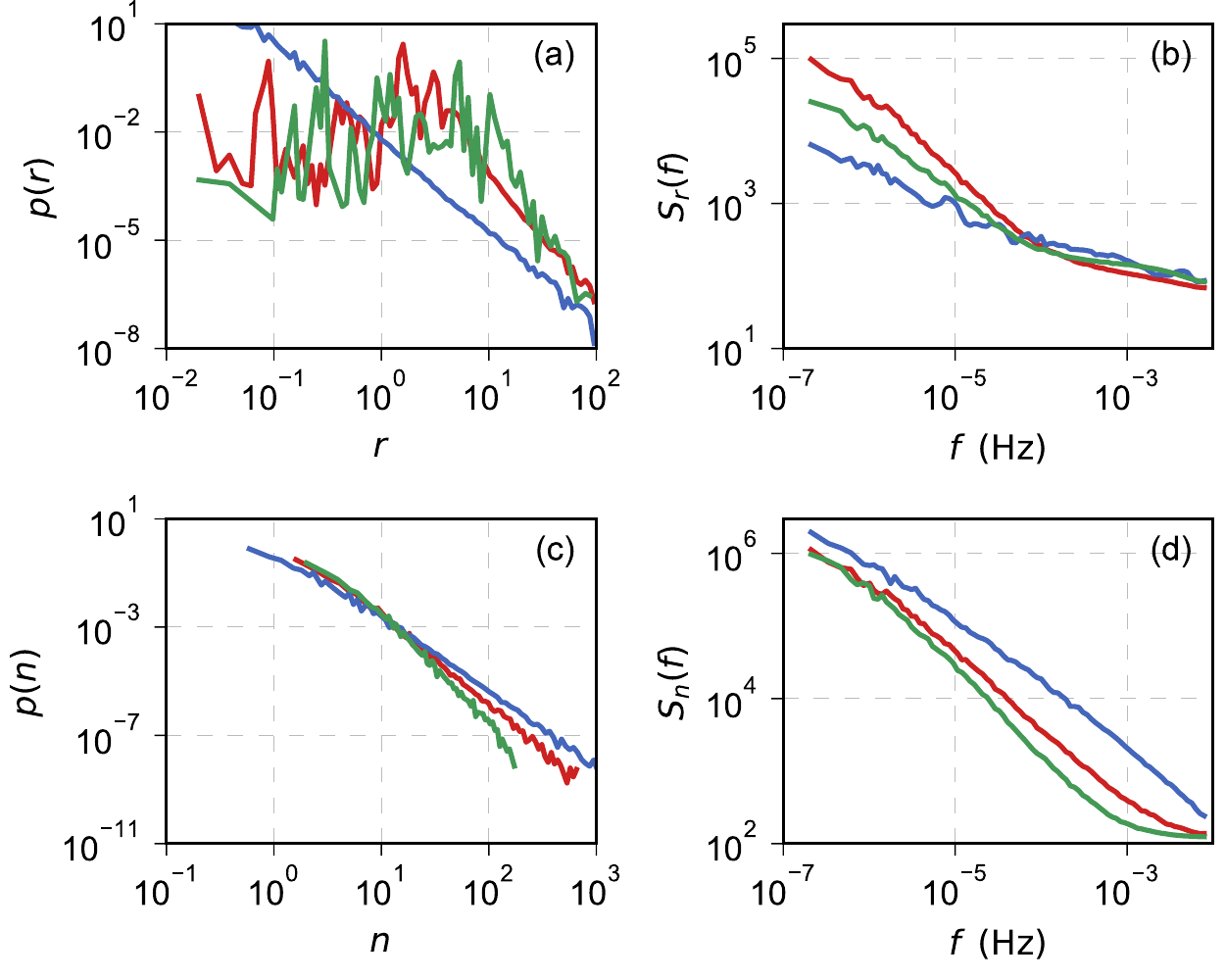}
\par\end{centering}
\caption{(color online) Influence of $\lambda_{tc}$ parameter on the model's
statistical properties. Parameter values are identical to the ones
used in Fig.~\ref{fig:model-btc} except: $\lambda_{tc}=250$ (blue
curve), $\lambda_{tc}=2.5$ (green curve).\label{fig:params-latc}}
\end{figure}

\begin{figure}
\begin{centering}
\includegraphics[width=0.7\textwidth]{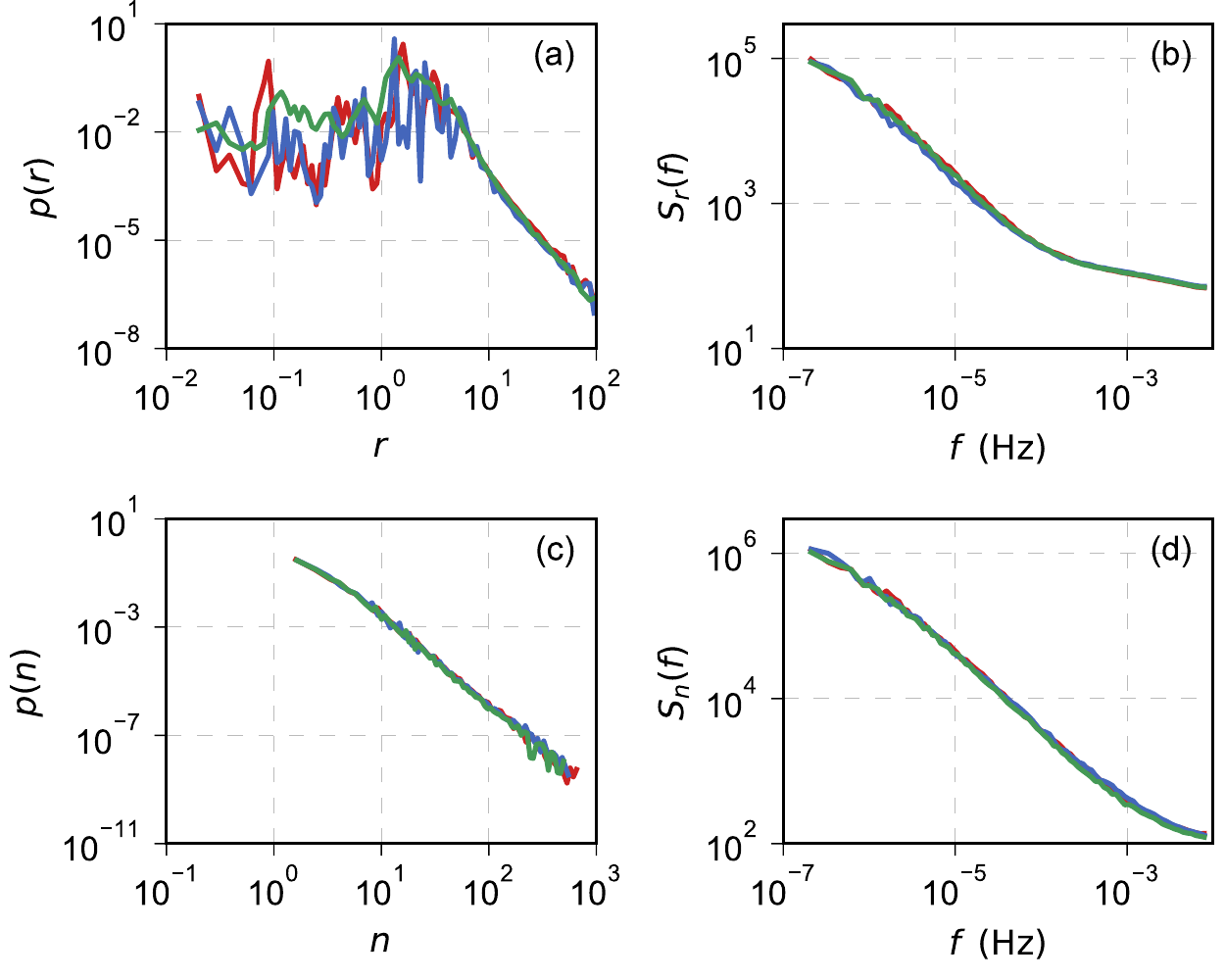}
\par\end{centering}
\caption{(color online) Influence of $\lambda_{tf}$ parameter on the model's
statistical properties. Parameter values are identical to the ones
used in Fig.~\ref{fig:model-btc} except: $\lambda_{tf}=750$ (blue
curve), $\lambda_{tc}=7.5$ (green curve).\label{fig:params-latf}}
\end{figure}

As we have seen in the previous section larger $\lambda_{tc}$ and
$\lambda_{tf}$ values force the realized prices to more closely follow
the equilibrium prices. While looking at the statistical properties
of the model we see that $\lambda_{tf}$ does not seem to have a noticeable
effect (see Fig.~\ref{fig:params-latf}). This is because the fundamentalists
activate only if the current price deviates far from the the fundamental
price and their trades rapidly push the price back to the fundamental
price. On the other hand $\lambda_{tc}$ seems to have a significant
effect (see Fig.~\ref{fig:params-latc}): larger values lead to the
fatter tails of the PDFs, while the PSDs flatten.

\begin{figure}
\begin{centering}
\includegraphics[width=0.7\textwidth]{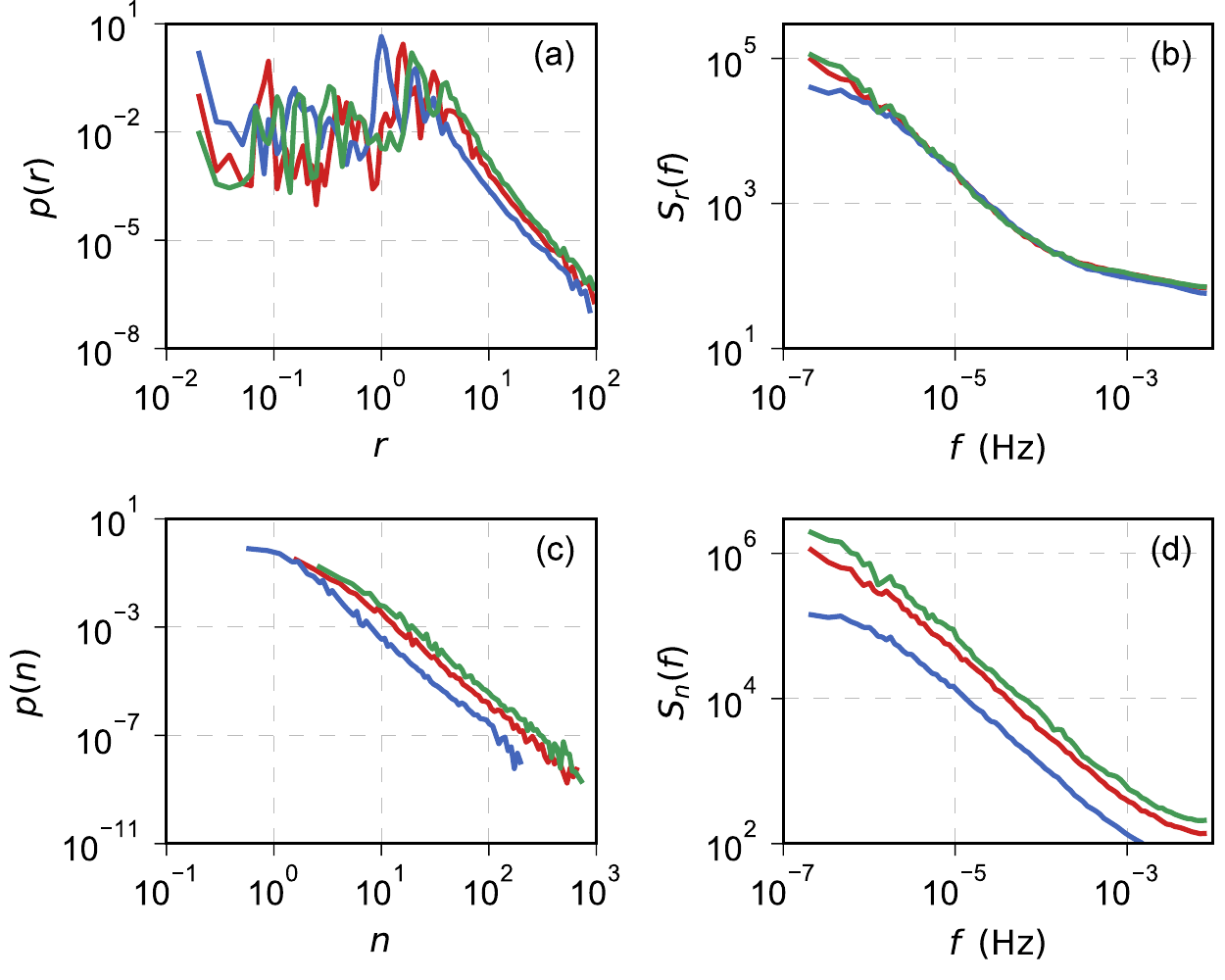}
\par\end{centering}
\caption{(color online) Influence of $\lambda_{0}$ parameter on the model's
statistical properties. Parameter values are identical to the ones
used in Fig.~\ref{fig:model-btc} except: $\lambda_{0}=4$ (blue
curve), $\lambda_{0}=0.04$ (green curve).\label{fig:params-la0}}
\end{figure}

As we can see in Fig.~\ref{fig:params-la0} changing $\lambda_{0}$
values does not have a significant effect besides increasing the overall
level of trading activity.

\begin{figure}
\begin{centering}
\includegraphics[width=0.7\textwidth]{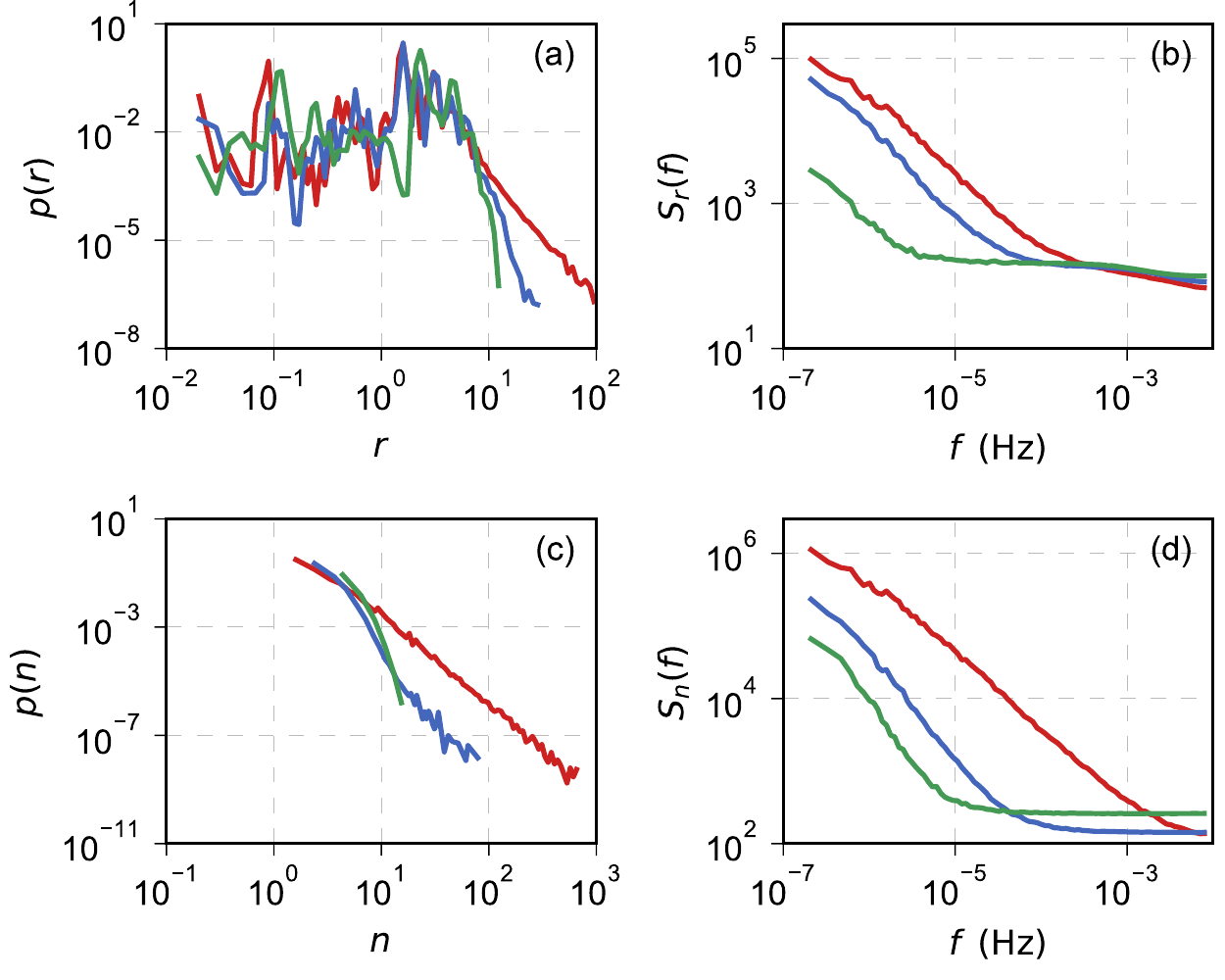}
\par\end{centering}
\caption{(color online) Influence of $\alpha$ parameter on the model's statistical
properties. Parameter values are identical to the ones used in Fig.~\ref{fig:model-btc}
except: $\alpha=1$ (blue curve), $\alpha=0$ (green curve).\label{fig:params-alpha}}
\end{figure}

Changing the power of the feedback scenario $\alpha$ seems to have
an adverse effect (see Fig.~\ref{fig:params-alpha}): with smaller
values the tails of the PDFs become lighter and the slopes of the
PSDs become steeper.

\begin{figure}
\begin{centering}
\includegraphics[width=0.7\textwidth]{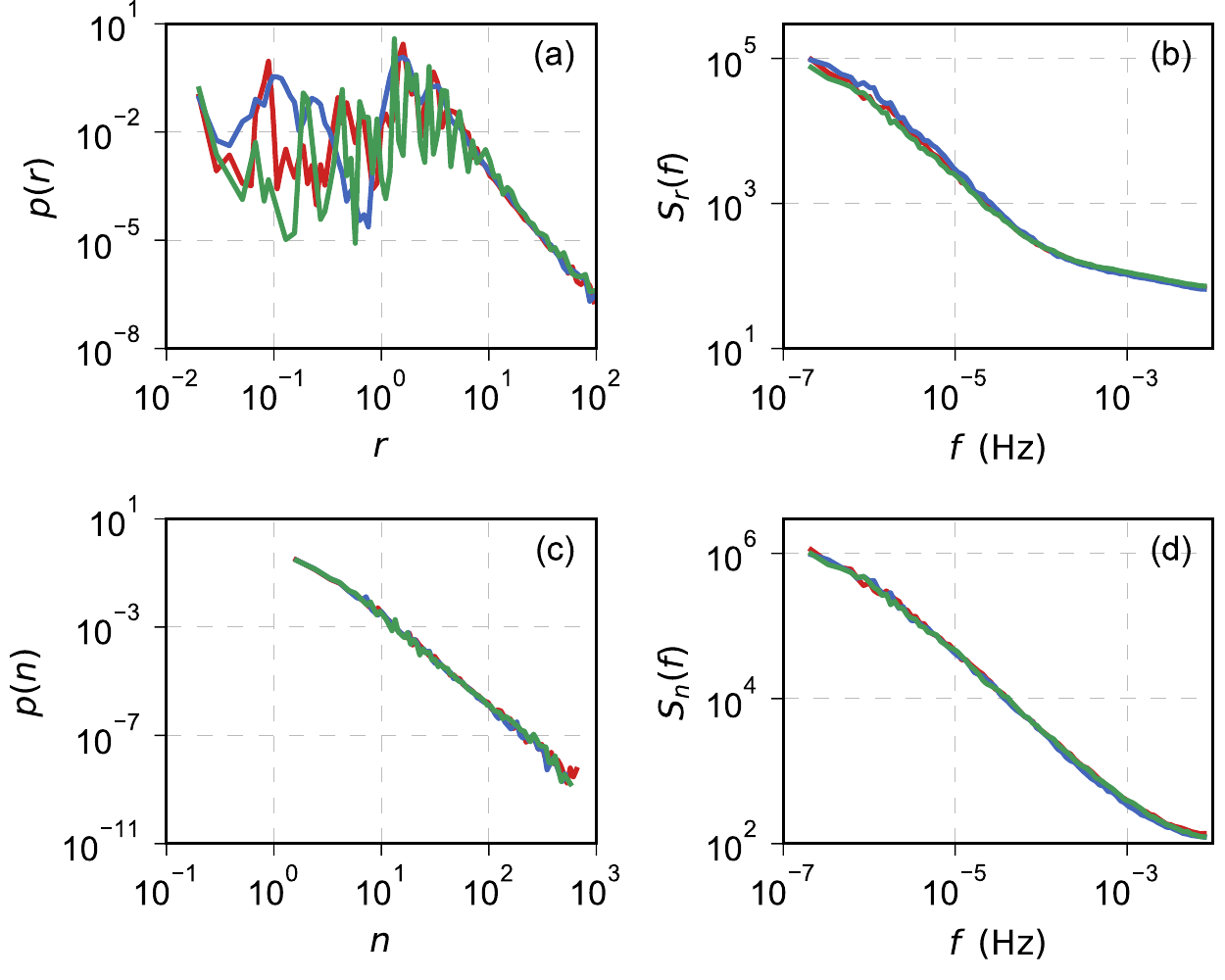}
\par\end{centering}
\caption{(color online) Influence of $k$ parameter on the model's statistical
properties. Parameter values are identical to the ones used in Fig.~\ref{fig:model-btc}
except: $k=16$ (blue curve), $k=1$ (green curve).\label{fig:params-k}}
\end{figure}

\begin{figure}
\begin{centering}
\includegraphics[width=0.7\textwidth]{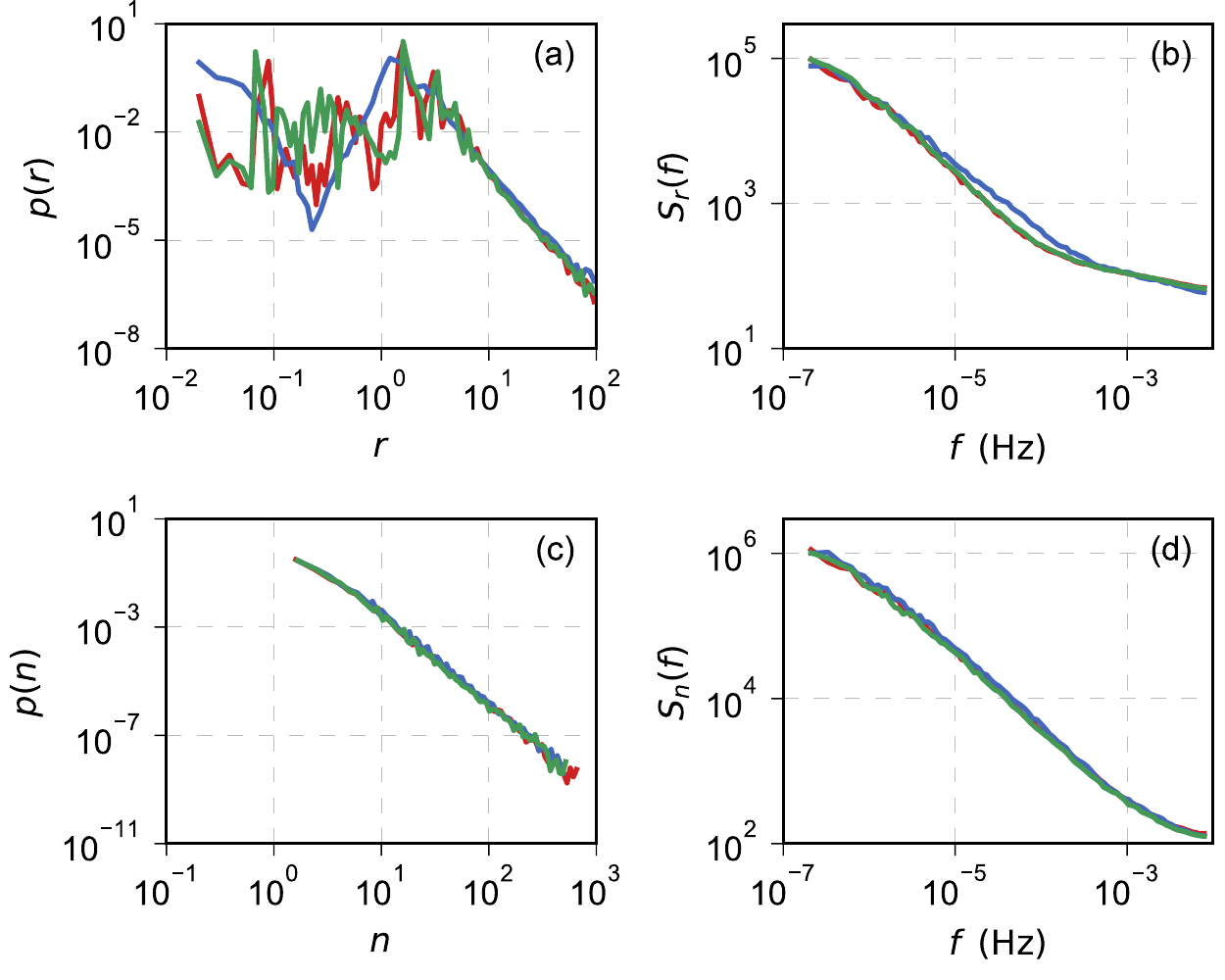}
\par\end{centering}
\caption{(color online) Influence of $\theta$ parameter on the model's statistical
properties. Parameter values are identical to the ones used in Fig.~\ref{fig:model-btc}
except: $\theta=100\,\mathrm{p.u.}$ (blue curve), $\theta=2.5\,\mathrm{p.u.}$
(green curve).\label{fig:params-theta}}
\end{figure}

Interestingly, as can be seen in Figs.~\ref{fig:params-k} and \ref{fig:params-theta},
the parameters influencing the overall shape of the order book itself,
$k$ and $\theta$, do not seem to have any effect on the statistical
properties obtained from the normalized time series. This is result
could be attributed to the simplifying assumption about the homogeneity
of the valuation. If valuation would be allowed to be heterogeneous
as in \cite{Kanazawa2018PRL,Kanazawa2018PRE}, the order book shape
parameters would likely have a more profound effect on the observed
statistical properties.

\section{Conclusions\label{sec:conclusions}}

Here we have proposed an order book model with herd behavior, which
is able to reproduce the main stylized facts of the financial markets.
The order book model with herd behavior was built upon empirical insights
by Kanazawa \textit{et al.} \cite{Kanazawa2018PRL,Kanazawa2018PRE},
who have studied a very detailed records of the order book level events,
and our previously proposed theoretical ABM \cite{Kononovicius2012PhysA,Gontis2014PlosOne},
which is known to successfully reproduce the statistical properties
of the high--frequency absolute return. Incorporating order book
dynamics improves upon our previous work in numerous ways. First of
all, we were able to scrap two not very realistic, but still common
in the literature, assumptions: we no longer need to introduce an
efficient market maker to define the market price (introduced in \cite{Kononovicius2012PhysA}),
we also no longer need to introduce the exogenous noise as was done
in \cite{Gontis2014PlosOne}. Another key improvement is that now
we are able to consider statistical properties of the trading activity
alongside the statistical properties of absolute return.

Using simulated annealing, optimizing the root mean squared error
of the worst match, we were able to calibrate the model parameters
to match the Bitcoin's statistical properties on one minute timescale.
Calibrating the model to match the statistical properties observed
in NYSE (one minute timescale) was not as successful, which indicates
that the model still lacks something. We believe that introducing
heterogeneity into chartist and fundamentalist valuation of the stock
might be the key, but this would further complicate the model introducing
additional parameters. Another possibility would be to complicate
the modeling of the chartist mood swings. Finally, the model structure
itself suggests that some of the parameter values could be gleaned
from the order book level data. Currently we are gathering publicly
available Bitcoin order book level data in hopes to use the collected
data to improve calibration of the model.

To limit the complexity of the model and to allow for analytical tractability
we have made few highly simplifying assumptions. Most restrictive
of them are homogeneous valuation, unit order volume and mood dynamics
assumptions. As relaxing these assumptions could enrich the dynamics
of the proposed order book model with herding behavior, we consider
doing so in the future works.

A relevant future goal could be improvement economic soundness of
the trading strategies and switching behavior of the agents. This
would make model more easily comparable with the other recent approaches
\cite{Kaizoji2015,Biondo2018Eco,Biondo2018JEcoIntCo,Biondo2018SEF,Navarro2017Plos,Llacay2018CompMatOrgT}
as well as open up the possibility to provide a deeper insight for
the economic policy makers. Similar transition was already undertaken
by Biondo (starting from \cite{Biondo2016JNTF} and arriving to \cite{Biondo2018Eco,Biondo2018JEcoIntCo,Biondo2018SEF}).

We could also extend our approach further by considering its dynamics
analytically. This should be possible, because one of the underlying
models can be alternatively described using stochastic differential
equations, while the order book part of the model can also be approached
analytically using Bogoliubov--Born--Green--Kirkwood--Yvon hierarchy.
Though fully integrating the both parts could prove to be a challenging
task. There could be a couple of possible approaches: a superstatistical
or a coupled SDE approach as used in \cite{Ruseckas2012ACS,Ruseckas2016JStatMech}
or a coarse--grained approximation of the model as discussed in \cite{Zou2012PRE}.

\section*{Acknowledgement}

This research was funded by the European Social Fund under the No
09.3.3--LMT--K--712 ``Development of Competences of Scientists,
other Researchers and Students through Practical Research Activities''
measure.

%\bibliographystyle{IEEEtran}
%\bibliography{library}

\end{document}